\documentclass[aps,pre,twocolumn,superscriptaddress,10pt]{revtex4-2}
\usepackage[utf8]{inputenc}
\usepackage{color}
\usepackage{amsmath}
\usepackage{amssymb}
\usepackage{xcolor} 
\usepackage{graphicx}
\usepackage{esint}
\usepackage{appendix}
\usepackage{comment}  
\usepackage{hyperref}
\usepackage{soul}
\usepackage{xcolor}
\setstcolor{red}
\makeatletter
\@ifundefined{textcolor}{}
{%
\definecolor{red}{rgb}{0.75, 0.1, 0.1}
 \definecolor{BLACK}{gray}{0}
 \definecolor{WHITE}{gray}{1}
 \definecolor{RED}{rgb}{1,0,0}
 \definecolor{GREEN}{rgb}{0,1,0}
 \definecolor{BLUE}{rgb}{0,0,1}
 \definecolor{CYAN}{cmyk}{1,0,0,0}
 \definecolor{MAGENTA}{cmyk}{0,1,0,0}
 \definecolor{YELLOW}{cmyk}{0,0,1,0}
}

\makeatother

\begin{document}

\title{Initial-State Typicality in Quantum Relaxation
}

\author{Ruicheng Bao}
\email{Contact author: ruicheng@g.ecc.u-tokyo.ac.jp}
\affiliation{Department of Physics, Graduate School of Science,
The University of Tokyo, Hongo, Bunkyo-ku, Tokyo 113-0033, Japan}

\begin{abstract}
Relaxation in open quantum systems is fundamental to quantum science and technologies. Yet, the influence of the initial state on relaxation remains a central, largely unanswered question. Here, by systematically characterizing the relaxation behavior of generic initial states, we uncover a typicality phenomenon in high-dimensional open quantum systems: relaxation becomes nearly initial-state-independent as system size increases under verifiable conditions. Crucially, we prove this typicality for many thermalization processes above a size-independent temperature. Our findings extend the typicality to transient open quantum dynamics, in turn identifying a class of systems where two widely used quantities---the Liouvillian gap and the maximal relaxation time---merit re-examination. We formalize this with two new concepts: the ``typical strong Mpemba effect'' and the ``typical relaxation time''. Beyond these conceptual advances, our results provide practical implications: a scalable route to accelerating relaxation and a typical mixing-time benchmark that complements conventional worst-case metrics for quantum simulations and state preparation. 

\end{abstract}
\maketitle

\textit{Introduction}---Relaxation is a ubiquitous phenomenon in nature and a central process in open quantum systems. The relaxation time, also known as the mixing time, is a key figure of merit in diverse areas, including quantum information and computing \cite{nielsen01quantum,24quantum_glauber,25arxiv_rapid_information}, quantum simulations \cite{11pra_simulations,13njp_simulations,14rmp_simulations,25nature_chen}, state preparation \cite{chen23_preparation,24prr_lin_prep,mi24stable_science,lin25diss_preparation,hu25prl_asy}, and quantum heat engines \cite{06prl_heat_engine,07pre_heat_engine,09pre_heat_engine,19pre_heat_engine,21prl_heat_engines}. Over the past decades, influential lines of work employing rigorous mixing-time \cite{diaconis1996cutoff,Kastoryano12_cutoff,cubitt15rapidmixing,23PRL_rapid_mixing,24arxiv_optimal_mixing,lin_25cmp_rapidmixing,25cmp_rapid_commute,25arxiv_KMS_rapid,25PRXQuantum_fast_mixing,25PRL_mixing} and spectral analysis \cite{lu2017nonequilibrium,klich2019mpemba,21prl_QMpemba,Goold_24prl,bao_25prl,Ueda_skin_effect,23prb_skin_effect,Mori20resolving,21PRR_mori,21prl_relax_rate_bound,23PRL_mori,24prl_mori,bao25_pre} have clarified how the structure of the dynamics controls relaxation in open systems. These approaches have been instrumental in characterizing remarkable phenomena such as the cutoff phenomenon in classical and quantum Markov processes \cite{diaconis1996cutoff,Kastoryano12_cutoff}, the Liouvillian skin effect \cite{Ueda_skin_effect,23prb_skin_effect} and the rapid mixing \cite{cubitt15rapidmixing,23PRL_rapid_mixing,24arxiv_optimal_mixing,lin_25cmp_rapidmixing,25cmp_rapid_commute,25arxiv_KMS_rapid,25PRXQuantum_fast_mixing}. A unifying aspect of these studies is their focus on identifying a single, initial-state-independent timescale to characterize a given dynamics. However, this approach overlooks the conceptually and practically important influence of the initial state on relaxation. 
Since every relaxation process begins from a specific state, its impact cannot be disregarded. 

A shift of perspective emerged with the Markovian Mpemba effect established in \cite{lu2017nonequilibrium,klich2019mpemba}: under fixed dynamics, different initial states can relax at dramatically different rates. The Mpemba effect has been intensively studied in classical systems \cite{20prl_raz_precooling,20nature_mpemba,23prl_raz_mpemba_boundary,22prl_mpemba_active,22JPCL_SL,23PRR_bao,23jcp_hasty_lu,ohga_mpemba,24PRXLife_lu_mpemba,25prl_tan_majormpemba,25prl_mpemba_pontus}, and inspired other studies on anomalous initial-state effect \cite{godec_20prl_asymmetry,21prr_asymmetry_tan,godec24_nphy}. Follow-up works extended the Mpemba effect to quantum systems \cite{21prl_QMpemba,22pra_qmpemba_carollo,Goold_24prl,23prl_dot_qmpemba,24prr_qmpemba_open,24prl_Qmpemba_open,25prl_nonmarkov_mpemba,bao_25prl,25prl_ergotropic_mpemba,25ncomm_observe_mpemba,23ncomm_entangle_asymmetry,24prl_qmpemba_closed,24prl_circuit_qmpemba,24prl_qmpemba_simulations,25prl_circuit_qmpemba,25prb_qmpemba_closed,25arxiv_resource_mpemba}, inaugurating a line of initial-state engineering for accelerating relaxation. See \cite{25review_mpemba,25review_qmpemba} for recent reviews. Despite these advances, a key question remains open: moving beyond the 
specific, pre-selected initial states and small systems common in Mpemba-effect studies, is there a universal relaxation behavior for generic initial states, how does it scale with system size, and---if affirmative---what scalable practical utility does it enable?
 
Here we address this question by proposing a theoretical framework for characterizing the universal relaxation behavior of generic initial states across different system sizes. Crucially, we uncover a typicality phenomenon (concentration) in the relaxation of high-dimensional open quantum systems. Typicality is of fundamental importance and is regarded as a cornerstone of statistical mechanics \cite{06nphy_concentration,06PRL_concentration}. To date, it has been established mainly for equilibrium quantum states \cite{06nphy_concentration,06PRL_concentration,06cmp_concentration}, closed quantum dynamics \cite{09PRE_concentration,09PRL_concentration,16ncomm_Reimann,tasaki16typicality}, quantum machine learning \cite{18ncomm_barren,24prxquantum_barren} and classical wave systems \cite{18PRL_Tian_concentration}, while its counterpart for transient open quantum dynamics has remained largely unknown. We show that, under a large class of conditions, the relaxation dynamics becomes insensitive to the initial state. We provide explicit examples where these conditions are satisfied, including spin chains, rapid-mixing systems and high-temperature thermalization. Insights from random matrix theory suggest the potential universality of this phenomenon. Moreover, the phenomenon prompts a re-evaluation of two well-established concepts: the Liouvillian gap and the maximal relaxation time.

In addition to these new fundamental insights on open quantum dynamics, our results yield practical utility: the typicality phenomenon implies that simple randomized preparations via techniques like the unitary 2-design \cite{09PRA_2design} may suffice to achieve fast relaxation. This offers a scalable route to accelerating quantum relaxation that avoids fine-tuning and detailed prior knowledge of the dynamics, thereby directly relevant to large-scale quantum platforms. Furthermore, our work provides a novel typical-case efficiency benchmark for dissipative quantum tasks, such as the evaluation of quantum Gibbs samplers \cite{25nature_chen}.

\textit{Setup---}We consider a Markovian open quantum system living in a Hilbert space $\mathcal{H}$ of dimension $d$, whose dynamics is described by the Lindblad-Gorini-Kossakowski-Sudarshan master equation
\begin{equation}
\frac{d\rho}{dt}=\mathcal{L}(\rho)\equiv-i[H,\rho]+\sum_{i}\left[J_{i}\rho J_{i}^{\dagger}-\frac{1}{2}\{J_{i}^{\dagger}J_{i},\rho\}\right].\label{Lindbladian}
\end{equation}
The superoperator $\mathcal{L}$ is referred to as Lindbladian throughout this work. In the above, $H$ is the Hamiltonian of the system and the jump operators
$J_{i}$ describe the dissipative coupling to the environment, $[A,B]:=AB-BA$ is the commutator and $\{A,B\}:=AB+BA$ denotes the anticommutator for any operators $A,\ B$. The eigenvalues of $\mathcal{L}$ are denoted $\lambda_{k}$ and
ordered such that $0=\lambda_1 \geq \text{Re}(\lambda_k)\geq \text{Re}(\lambda_{k+1}),\ k\geq2$. 
Assuming that $\mathcal{L}$ is diagonalizable,
we define $R_{k}$ and $L_{k}$ as the right and left
eigenmatrices corresponding to $\lambda_k$, satisfying $\mathcal{L}(R_{k})=\lambda_{k}R_{k}$ and
$\mathcal{L^{\dagger}}(L_{k})=\lambda_{k}^{*}L_{k},\ k=1,...,d^{2}$,
with the dual superoperator $\mathcal{L}^{\dagger}$ governing the Heisenberg-picture
dynamics of observables:
\begin{equation*}
    \mathcal{L}^{\dagger}(O)=i[H,O]+\sum_{i}\left[J_{i}^{\dagger}OJ_{i}-\frac{1}{2}\{J_{i}^{\dagger}J_{i},O\}\right].
\end{equation*}
We normalize the eigenmatrices by the biorthogonal condition
\begin{equation}
{\rm tr}(L_k^{\dagger}R_h)\equiv\delta_{kh},\quad \|R_k\|_{1}=1,\label{normalized}
\end{equation}
where $\|A\|_{1}:={\rm tr}(\sqrt{A^{\dagger}A})$ is the trace norm. The additional constraint $\|R_k\|_{1}=1$ excludes any size-dependence from $R_k$ on the relaxation. 
We assume $\mathcal{L}$ has a unique stationary state $\rho_{\rm ss}$ (however, our analysis can be directly extended to cases where multistability exists \cite{16prx_jiang_geometry}). Then, given an arbitrary initial state $\rho_{0}$, the system density matrix at time $t$
is
\begin{equation}
\rho(t)=e^{t\mathcal{L}}(\rho_{0})=\rho_{\rm ss}+\sum_{k=2}^{d^{2}}e^{\lambda_{k}t}a_{k}R_{k},
\end{equation}
where coefficients $a_{k}\equiv {\rm tr}(L_{k}^{\dagger}\rho_{0})$ determine the overlaps between the $k$th relaxation mode and the initial state. 

In the long-time limit, the density matrix evolves as
\begin{equation}\label{long_time}
\|\rho(t)-\rho_{\rm ss}\|_{1}\overset{\text{large } t}{\sim}
\begin{cases}
|e^{\lambda_2 t}a_2|, & \text{real $\lambda_2$},\\[6pt]
|e^{\lambda_2 t}a_2|+|e^{\lambda_2^* t}a_2^*|, & \text{others},
\end{cases}
\end{equation}
where we use $\|R_2\|_1=\|R_2^{\dagger}\|_{1}=1$.
From Eq. \eqref{long_time}, it is clear that the relaxation timescale is determined by the slowest mode, i.e., $\tau_{\rm rel}\sim 1/|{\rm Re}(\lambda_2)|$. Eq. \eqref{long_time} also implies that, the relaxation starting from an initial state with a smaller $|a_2|$ will be faster than that from one with a larger $|a_2|$ \footnote{This can be more clearly seen by defining an $\epsilon$-relaxation time $\tau^{\epsilon}_{\rm rel}$ as in \cite{Ueda_skin_effect}. It is defined by the relation $|a_2|e^{-|{\rm Re}(\lambda_2)| \tau^{\epsilon}_{\rm rel}}=\epsilon$, which gives $\tau^{\epsilon}_{\rm rel}=\frac{1}{|{\rm Re}(\lambda_2)|}\log(|a_2|/\epsilon)$.}. In particular, if $a_2=0$, the relaxation timescale becomes $1/|{\rm Re}(\lambda_3)|$, termed the strong Mpemba effect (SME) \cite{klich2019mpemba}. While the slowest mode, which has been the sole focus of most previous studies, governs the long-time dynamics, at intermediate times other modes $(k>2)$ can also contribute, motivating a more complete view.

The existing literature has largely focused on specific, engineered initial states, such as those tailored to produce the SME \cite{21prl_QMpemba,Goold_24prl}. This leaves a more fundamental question unanswered: What is the relaxation behavior for a \textit{generic} initial state chosen from the Hilbert space? Answering this requires a shift from analyzing specific values of $a_k$ to understanding their collective statistical properties.

\textit{Typical relaxation behaviors with respect to the initial state---}We now address this question under a fixed $\mathcal{L}$. To make the notion of a ``generic state'' concrete, we model it by considering states $\rho_0$ drawn from a few representative random ensembles and analyzing the resulting statistics of their coefficients $a_k(\rho_0)$. In this setting, the overlaps $a_k$ become random variables, while $L_k,\ R_k$ and $\lambda_k$ remain fixed, allowing us to characterize the typical initial-state dependence of relaxation. For each ensemble, we focus on the first two moments of this distribution due to their experimental feasibility.  

We begin by examining an ensemble generated by an operationally convenient procedure. We start with any convenient initial state $\rho'_0$ and apply a random unitary (from a unitary $2$-design) to prepare the initial state for the relaxation, denoted as $\rho_0=U\rho_0'U^{\dagger}$. When $\rho'_0$ is a pure state, the situation reduces to having the initial state of the relaxation process of interest randomly sampled from a distribution that precisely matches the first two moments of the Haar measure. The variance of $a_k={\rm tr}(L_k^{\dagger}\rho_0)$ over the random unitary $U$ in a 2-design is given by (see Supplemental Material \cite{supplemental_material} for the derivation)
\begin{equation}\label{var_Haar}
    {\rm Var}_U(a_k)=\frac{[{\rm tr}(\rho_0'^{2})-\frac{1}{d}]\left[\|L_k\|_{2}^2-\frac{|{\rm tr}(L_k^{\dagger})|^2}{d}\right]}{d^2-1},
\end{equation}
where the variance is defined as ${\rm Var}(a_k):=\langle |a_k|^2\rangle-|\langle a_k\rangle|^2$ and $\|A\|_{2}:=\sqrt{{\rm tr}(A^{\dagger}A)}$ is the Hilbert-Schmidt (HS) norm. The inequalities $|{\rm tr}(L_k^{\dagger})|^2\leq \|\mathbb{I}_d\|^2_{2}\|L_k\|_{2}^2=d \|L_k\|_{2}^2$ and ${\rm Tr}(\rho_0^2)\geq1/d$ guarantee that the variance is non-negative. To proceed, we introduce the diagonal
eigenvalue condition number \cite{trefethen20spectra,18cmp_overlap}  $O_k:=\|L_k\|_{2}\|R_k\|_{2}$, which controls the sensitivity of $\lambda_k$ to perturbations \cite{trefethen20spectra}. The HS norm and the trace norm are connected via the relation $\|A\|_2\leq\|A\|_1\leq \sqrt{d}\|A\|_2$. Consequently, the variance is upper bounded as ($\|R_k\|_1=1$)
\begin{equation}
    {\rm Var}_U(a_k)< \frac{\|L_k\|_{2}^2}{d^2}\leq \frac{O_k^2}{d}.
\end{equation}
Thus, if $O_k$ grows slower than $\sqrt{d}$, the variance of $a_k$ converges to 0 in the $d\to \infty$ limit, e.g. $O_k\sim d^{a},\ a\in [0,1/2)$ or grows logarithmatically. That is, the variance exponentially decays to zero as the system size increases (e.g., for a one-dimensional quantum chain consisting of $n$ qubits, $d=2^n$) and $a_k$ concentrates sharply around its mean value,
\begin{equation}\label{mean_value}
    \langle a_k\rangle=\frac{{\rm tr}(L_k^{\dagger})}{d}.
\end{equation}
That is, most initial states have the same $a_k\approx \langle a_k\rangle$, though they are not necessarily close to each other.
This establishes a sufficient condition of concentration of the $k$th relaxation mode. 

Notably, $O_k$ has a clear physical interpretation: it measures the robustness of the $k$th mode's characteristic timescale against perturbations on the Lindbladian. A large $O_k$ signifies that the mode is physically fragile, as its timescale $1/|{\rm Re}(\lambda_k)|$ is highly sensitive to small uncertainties or perturbations within the system's Hamiltonian or dissipative channels, see \footnote{For a perturbation $\delta\mathcal{L}$ to the Lindbladian, the resulting first-order change in the eigenvalue $\lambda_k$ is given by $\delta\lambda_k = \text{Tr}(L_k^\dagger \delta\mathcal{L} R_k)$. This leads to the bound $|\delta\lambda_k| \le O_k \|\delta\mathcal{L}\|_{\rm op}$, where $\|\cdot\|_{\rm op}$ is the operator norm. See \cite{supplemental_material,trefethen20spectra} for more details}. Thus, relaxation modes that are more robust to perturbations are easier to concentrate. 

The random mixed state generated by performing a random unitary on a mixed state, may be appealing for further experimental verifications and applications due to its relative operational simplicity. However, such mixed states are not fully random, because unitary operations do not affect the eigenvalues of a density matrix, so their spectrum is not random. We thus need to consider random mixed states induced by the partial trace to ensure true randomness. For an initial state $\rho_0$ randomly selected from a truly random mixed state ensemble (the HS measure \cite{01_induced_measure}), the variance scales as \cite{supplemental_material}
\begin{equation}\label{var_HS}
    {\rm Var}(a_k)=\frac{\|L_k\|_2^2-\frac{|{\rm tr}(L_k^{\dagger})|^2}{d}}{d(d^2+1)}< \frac{\|L_k\|_2^2}{d^3}\leq \frac{O_k^2}{d^2},
\end{equation}
while the average $\langle a_k\rangle$ is still given by Eq. \eqref{mean_value}. Thus, for a random mixed state, the sufficient condition for concentration weakens to $O_k$ grows slower than $d$.

A more meticulous description of the concentration phenomenon is the probability of deviating from mean value. Using Chebyshev's inequality, we obtain
\begin{equation}
    P(|a_k-\langle a_k\rangle|>\epsilon)\leq \frac{{\rm Var}(a_k)}{\epsilon^2}<\frac{\|L_k\|_2^2}{d^{q}\epsilon^2}\leq\frac{O_k^2}{d^{q-1}\epsilon^2},
\end{equation}
where $q=2$ for random pure states and $q=3$ for random mixed states. This offers a more direct intuition of concentration: in the thermodynamic limit, most initial states yield a value of $a_k$ that lives within a small region around the mean. 

On the other hand, if $\|L_k\|_2$ grows faster than $\mathcal{O}(d^{3/2})$, the variance will diverge in the $d\to \infty$ limit for both pure and mixed random initial states. There are known examples where this is the case: the Liouvillian skin effect \cite{Ueda_skin_effect} and the super-exponential overlap coefficients \cite{Mori20resolving}. This drastic change may be termed the ``typicality transition''. 

So far, we require a ``generic'' initial state not to have any preference (uniformly random), thereby faithfully reflecting the underlying properties of $\mathcal{L}$. However, we demonstrate in \cite{supplemental_material} that the concentration can still occur with an arbitrary physical constraint (or, prior knowledge) on the initial state. 

The scaling conditions for the appearance of concentration of quantum relaxation modes or divergence of their variances are a main prediction of this work. If the concentration conditions are satisfied, the initial state becomes irrelevant for relaxation dynamics as the dimension $d$ increases. To be more specific, when $a_2$ concentrates, the long-time relaxation dynamics concentrates, and the relaxation time becomes irrelevant to the initial state; if other $a_k$ also concentrate, the full relaxation dynamics concentrates and even the early and intermediate-time dynamics are nearly independent of the initial state. This novel concentration phenomenon in the transient dynamics of open quantum systems is a complementary finding to results on the typicality in equilibrium states \cite{06nphy_concentration,06cmp_concentration,06PRL_concentration} and closed quantum dynamics \cite{16ncomm_Reimann,tasaki16typicality}. 

To provide a concrete understanding of the concentration phenomenon, we now present explicit examples demonstrating how the dimensional scaling of $O_k$ or $\|L_k\|_2$ meets the established requirement for concentration. We defer the discussion of some important implications and practical applications to the end of this work.

\textit{Example 1: Non-interacting dissipative $N$-qubit chain---}We consider a non-interacting system consisting of $N$ damped qubits, with Hamiltonian $H=\sum_{i=1}^NE\sigma^{z}_{i}/2$ and local jump operators $J_{i,\downarrow}=\sqrt{\gamma_1}\,\sigma_i^-,
\quad
J_{i,\uparrow}  =\sqrt{\gamma_0}\;\sigma_i^+$. Here, $\sigma^{z}_i$ are Pauli operators of the $i$th particle, $\sigma_i^-=(|0\rangle\langle 1|)_i$ and $\sigma_i^+=(|1\rangle \langle 0|)_i$. The system can be analytically solved to illustrate the concentration phenomenon. In this case, it can be verified that \cite{supplemental_material} $\|L_k\|_{2}\sim \mathcal{O}(2^{N/2})\sim\mathcal{O}(\sqrt{d})$. Thus, the variances for both ensembles exponentially converge to $0$ as ${\rm Var}_U(a_2)\sim d^{-1}\sim 2^{-N},\ {\rm Var}(a_2)\sim d^{-2}\sim 4^{-N}$. One can further add the dephasing noise $\sigma_i^z$, which will not change the scaling.

\textit{Example 2: Dissipative transverse-field Ising model (TFIM)---}We then numerically study an open TFIM of length $N$ with local dissipation. The Hamiltonian is $H =-J\sum_{i=1}^{N-1}\sigma_i^z\sigma_{i+1}^z
     -g\sum_{i=1}^{N}\sigma_i^x$,
and jump operators are the same as Example 1 (with a physical constraint $\gamma_1/\gamma_0= e^{\beta E},$ where $\beta=1/k_B T$ is the inverse temperature of multiple identical heat baths independently coupled to each qubit). Notably, $\rho_{\rm ss}$ may not be an equilibrium state in this example. The steady state is unique in both this model and Example 1, as the jump operators satisfy the sufficient condition for uniqueness \cite{24prl_qmpemba_simulations}. We find that the concentration of $a_2$ is robust in this model, across different parameter regimes ($J,\ g,\ \beta$), the details of a specific case are shown in End Matter. The robustness is reminiscent of the fact that there is no thermal phase transition in one-dimensional spin systems and there may be a connection to be explored.  

\textit{Sufficiently rapid relaxation may imply mode concentration---}Rapid mixing, a well-known phenomenon stating that the relaxation timescale for a system with size $n$ is of the order $\mathcal{O}(\log n)$, can imply the concentration of $a_2$. We prove that $\|L_k\|_2\leq 2\sqrt{d}\max_{\rho_0}|a_k|$ for any $k$. This inequality yields upper bounds for the variance: ${\rm Var}_U(a_k)\leq4(\max_{\rho_0}|a_k|)^2d^{-1} $ and ${\rm Var}(a_k)\leq 4(\max_{\rho_0}|a_k|)^2d^{-2}$. Indeed, rapid mixing implies $\max_{\rho_0}|a_2|$ is at most of the order $\mathcal{O}({\rm poly}(\log d))$ under an assumption used in \cite{Mori20resolving}, ensuring the variances vanish as $d\to \infty$. The rapid mixing has been proven for diverse systems \cite{cubitt15rapidmixing,23PRL_rapid_mixing,24arxiv_optimal_mixing,lin_25cmp_rapidmixing,25arxiv_KMS_rapid,25arxiv_lin_rapid_ground}. Interestingly, rapid mixing implies stability of relaxation \cite{cubitt15rapidmixing}, and concentration also implies more stable relaxation modes, which indicates deep connections between them. 

However, fast relaxation is not a necessary condition for typicality. Since the closing of Liouvillian gap is sufficient for a slow relaxation, the slow relaxation alone cannot imply $O_k$ grows fast. 

\textit{High-temperature thermalization is sufficient for initial-state typicality---}Importantly, we prove that for a large class of thermalization processes with a sufficiently high environment temperature, \textit{all} relaxation modes concentrate. To be specific, for the Lindbladian \eqref{Lindbladian} obeying the quantum detailed balance condition  (QDBC) with respect to a full-rank Gibbs state $\rho_{\beta}:=e^{-\beta H_0}/{\rm tr}(e^{-\beta H_0})$ with a Hamiltonian $H_0$ and an inverse temperature $\beta \equiv 1/(k_B T)$, we prove an upper bound for $O_k$ corresponding to any $\lambda_k$ \cite{supplemental_material}: \begin{equation}
    O_k\leq \sqrt{\frac{p_{\rm max}(\rho_{\beta})}{p_{\rm min}(\rho_{\beta})}}\equiv e^{\beta\Delta E/2},
\end{equation}  
where $p_{\max/\min}(\rho_{\beta})$ are maximum/minimum eigenvalues of $\rho_{\beta}$ and $\Delta E:=E_{\rm max}-E_{\rm min}$ is the energy range of $H_0$. Here, by QDBC we refer to the Davies map \cite{davies74markovian,davies79generators}  or the Kubo–Martin–Schwinger (KMS) condition \cite{lin_25cmp_rapidmixing} with respect to $\rho_{\beta}$ \footnote{See \cite{supplemental_material} for details of these conditions. It has been shown that, if a Lindbladian satisfies these conditions, $\rho_{\beta}=e^{-\beta H_0}/{\rm tr}(e^{-\beta H_0})$ is its steady state. Note that $H_0$ may not coincide with the system Hamiltonian $H$ of the Lindbladian, there could be a shift term, e.g., the Lamb shift \cite{25arxiv_KMS_rapid}.}. Lindbladians satisfying the KMS condition have been constructed \cite{chen23_preparation,chen23efficient} and have proven critical to quantum simulations \cite{25nature_chen}. Notably, we assume that $\rho_\beta$ is the unique steady state, which is natural in the context of thermalization, although the upper bound remains valid in the presence of multistability.

For most Hamiltonians without long-range interactions, $\Delta E\sim\mathcal{O}(n)\equiv cn$, where $c$ is a size-independent constant and $n$ is the system size. This requirement reflects the extensivity condition, which can also be fulfilled by long-range Hamiltonians when the Kac prescription is applied \cite{23rmp_long_range}. Then, take the many-qubit system as an example, if the temperature satisfies the condition $e^{\beta c}\leq 2\Rightarrow \beta <\beta_c:=\log2/c$, i.e., higher than a threshold irrelevant to the system size, $O_k$ has a smaller scaling than $\mathcal{O}(\sqrt{d})$ and all $a_k$ will concentrate around their mean for an initial Haar ensemble. For the HS ensemble, the threshold turns to $\beta_c'=2\log2/c$. 


\textit{Two important implications of the typicality---}First, the Liouvillian gap $1/|{\rm Re}(\lambda_2)|$ may not correctly quantify the relaxation time for a class of systems satisfying two conditions: (i) $a_2$ concentrates (ii) $|\langle a_2 \rangle| = |{\rm tr}(L_2^{\dagger})|/d$, vanishes. The second condition is met, for example, if ${\rm tr}(L_2)=0$ [e.g., for a complex $\lambda_2$ of a Davies map \cite{Goold_24prl}]. When both conditions hold, $a_2$ concentrates sharply around $0$. Our Example 1 also falls into this class. We term this the ``typical SME'' (TSME), as it implies the SME ($a_2\approx0$) is typical for generic initial states. In this regime, the relevant relaxation timescale may be set by $1/|{\rm Re}(\lambda_k)|$ with $k\geq3$, instead of $1/|{\rm Re}(\lambda_2)|$. We discuss the conditions for this phenomenon in detail in \cite{supplemental_material}. The TSME should be distinguished from the conventional SME, as it represents the intrinsic property of the underlying dynamics (Lindbladian). In contrast, the SME only reflects the property of a specific initial state.

Second, for systems with initial-state typicality, the relevance of the commonly adopted initial-state-independent maximal relaxation time (maximized over all initial states) is called into question, since it may deviate greatly from the \textit{typical} relaxation timescale and thus may not be physically representative. In our Example 1, all $a_k$ concentrate sharply around their mean, and $\max_{k}\langle a_k\rangle\sim \mathcal{O}(1)$. By contrast, the maximal relaxation time \cite{Mori20resolving,21PRR_mori} is calculated from $\max_{k, \rho_0}|a_k(\rho_0)|$, which can scale as $\mathcal{O}(d)$.

\textit{Definition: typical relaxation time---}We therefore propose the ``typical relaxation time'', the relaxation time from representative initializations whose $a_2$ concentrate around $\langle a_2\rangle$. As an illustrative definition analogous to that used for the maximal relaxation time \cite{Ueda_skin_effect,Mori20resolving,21PRR_mori}, one can define a typical timescale $\tau^{\epsilon}_{\rm rel}$ via the mean overlap of the slowest mode:
\begin{equation}
|\langle a_2\rangle| e^{-|{\rm Re}(\lambda_2)| \tau^{\epsilon}_{\rm rel}}=\epsilon.
\end{equation}
An alternative and more rigorous definition using the trace distance (the ``typical mixing time'') is given in the End Matter. We posit that this typical relaxation time is a more physically relevant quantity in the typicality regime. Notably, it still characterizes the intrinsic properties of the generator $\mathcal{L}$, but it does so by reflecting the statistically dominant behavior rather than an exponentially rare, worst-case scenario.


\textit{Potential practical utility: scalable acceleration of relaxation}---A key practical utility of initial-state typicality is enabling scalable relaxation acceleration. If a system is initially in a slow-relaxing state, applying a randomization protocol (such as a unitary 2-design) can kick it into the fast-relaxing typical set. The presence of a TSME, ($\langle a_2\rangle\to0$ and ${\rm Var}(a_2)\to0$ so that $a_2\approx 0$ for most states), is a powerful diagnostic indicating that this typical set is indeed characterized by rapid relaxation.

This randomization can be realized via a unitary 2-design, which only requires modest polynomial resources \cite{18ncomm_barren}. This can be applied directly to the system of interest, or used to generate a random mixed state by combining an auxiliary system with the system, performing the random unitary on the composite system, and then tracing out the auxiliary part. The technique has been performed experimentally in quantum many-body systems \cite{19prl_2design}, and its key component, the Clifford circuit, has been applied in systems up to 40 qubits \cite{25PRL_clifford}. For our purpose, an approximate $2$-design, implementable via scalable random circuits \cite{09cmp_random_circuits,arute19nature_supremacy,23nature_random}, is sufficient \cite{supplemental_material}. Remarkably, recent advances indicate that achieving an approximate $2$-design on $n$ qubits via random circuits can be done in circuit depth $\mathcal{O}(\log\log n)$ with ancillas \cite{25_huang_design,25_huang_strong,zhang25designs}, and in $\mathcal{O}(\log n)$ without ancillas \cite{25science_huang_random}. Additionally, suitable hybrid quantum–classical approaches may drive atypical states into the typical-relaxation regime in constant parallel depth when $\mathcal{O}(n)$ ancillas are available \footnote{Concretely, when $\mathcal{O}(n)$ ancilla qubits and intermediate measurements are available, the depolarizing action $\rho_0 \mapsto (1-p)\rho_0 + p\,\mathbb{I}/d$ can be realized in constant parallel depth per circuit via a hybrid sampling scheme, demonstrated on superconducting hardware up to 27 qubits \cite{24prr_hybrid_simulation}. Setting $p=1$ yields $a_2(\rho_0)=a_2(\mathbb{I}/d)=\langle a_2\rangle$ exactly. Our typicality result guarantees that this value is representative of almost all initial states; thus the unitary 2-design route and this shortcut are equivalent in the figure of merit $a_2$. Moreover, initial-state typicality provides an additional advantage in this case: since $\mathbb{I}/d$ lies at the center of the typical set, small imperfections in its preparation still correspond to states within that set, making the protocol intrinsically robust.}.

Our findings also imply a challenge for protocols that rely on finding atypical states. If $a_2$ concentrates to a large $\langle a_2\rangle$, then using random searching methods \cite{Goold_24prl} to find a rare $a_2\approx 0$ state may be inefficient, as the chance to escape from the typical set is exponentially small for many-body systems \cite{18ncomm_barren} . Thus, the search may need to be constrained in the weak-entanglement regime. 

Finally, our framework offers a representative benchmark for assessing the efficiency of quantum Gibbs samplers \cite{chen23_preparation,chen23efficient,25nature_chen}, complementing conventional worst-case metrics (see End Matter).

\textit{Conclusion and Outlook---}We established the phenomenon of initial-state typicality in the relaxation of open quantum systems. This finding has profound conceptual and practical implications. Conceptually, it challenges the universality of the Liouvillian gap and the maximal relaxation time as faithful descriptors of relaxation. To address this, we introduced the TSME as a diagnostic and proposed the typical relaxation time as a more relevant timescale in the typicality regime. On a practical level, the phenomenon underpins a scalable protocol for relaxation acceleration and offers new ways to assess the efficiency of algorithms like quantum Gibbs samplers.

Our work opens several promising avenues for future research. A concrete next step is to construct explicit physical models that exhibit a sharp transition into the typicality regime. 
The potential universality of the typicality itself, hinted at by random matrix theory, also warrants deeper investigation (see End Matter). Additionally, the concentration of slow modes implies that associated metastable states \cite{16prl_metastability} also exhibit typicality, potentially impacting quantum computation that leverages metastability \cite{allcock2021metastable,25PRA_metastability}. It is also interesting to extend our framework to broader settings. For instance, adapting this framework to bosonic systems with infinite-dimensional Hilbert spaces is not yet straightforward and would likely require a reformulation of the key ingredients (e.g., random ensembles) via physically motivated truncations \cite{22Quantum_truncation,24prx_boson_design,25lindbladian_truncation}. Perhaps most importantly, whether the typical relaxation time can help relax some of the conditions required for proving rapid mixing deserves further investigation, as it may provide alternative performance benchmarks in applications ranging from state preparation to quantum simulation \footnote{Specifically, a large separation between typical-case and worst-case mixing is expected when the maximal overlap with the slowest mode is large while the typical overlap is suppressed, a scenario exemplified by systems exhibiting the TSME. If the TSME can be established for specific low-temperature thermalization (e.g., in one-dimensional spin systems as suggested by the numerical results in Example 2), it may uncover concrete examples of typical rapid mixing even where conventional rapid mixing fails.}.

\textit{Acknowledgments---}I am grateful to Zongping Gong, Guohua Xu and Hongchao Li for useful discussions, and Yue Liu for a critical reading of the manuscript. I also acknowledge László Erdös for the kind help in private communications.  
R.~B. is supported by JSPS KAKENHI Grant No.\ 25KJ0766.

\textit{Data availability statement---}The code used to generate the data in Fig. \ref{2} is open available at \footnote{https://github.com/Ruicheng-12/Initial-state-typicality-in-quantum-relaxation}. 

\bibliography{refs}

\section*{End Matter}
\begin{figure}
    \centering
     \includegraphics[width=0.8\linewidth]{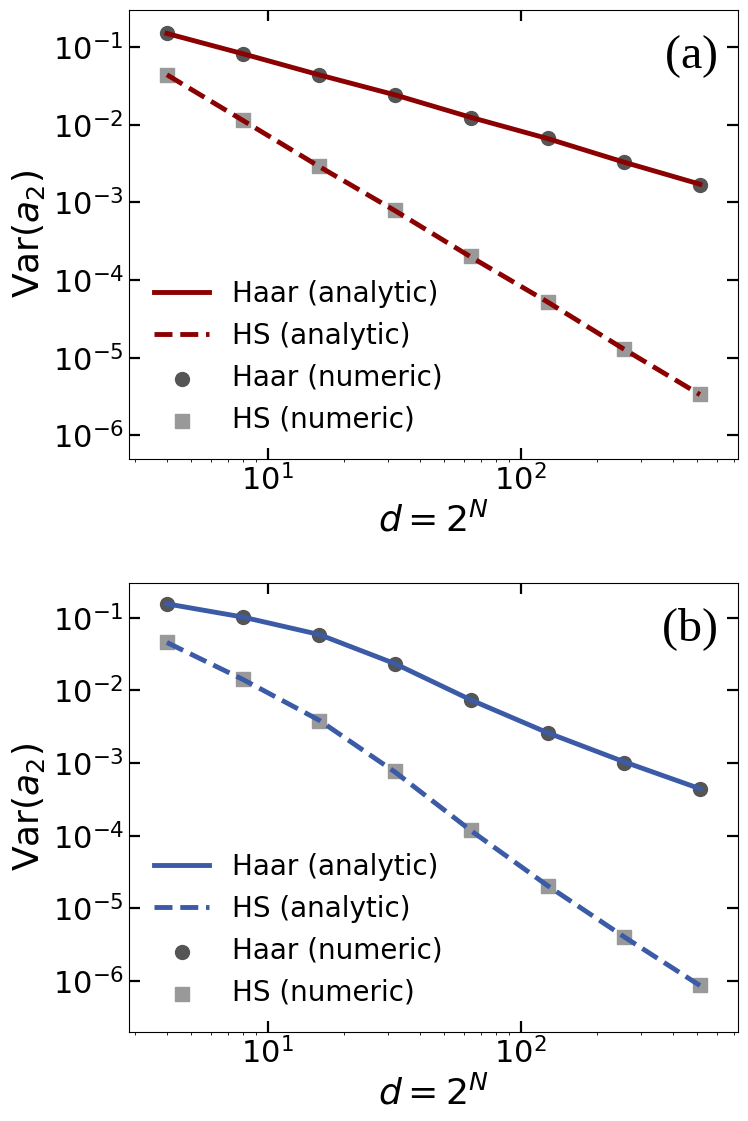}
     \caption{Variance of $a_2$ for the dissipative TFIM with different spin number $N$. The system size varies from $N=2$ to $N=9$ ($d=512$). Other parameters are $E=J=g=1.0$, $\gamma=0.5$. (a) $\beta=0.1$, high temperature (b) $\beta=100$, low temperature. Solid and dashed lines denote the analytical results obtained from Eqs. \eqref{var_Haar} and \eqref{var_HS}, respectively. Circle and square dots are purely numerical results obtained by random sampling $10000$ initial states from the Haar and the HS measure.}
    \label{2}
\end{figure}

\textit{Numerical results on the Example 2---}From the numerical results, the variance of $a_2$ in the interacting case converges at a rate similar to the non-interacting case derived above. For the high-temperature case ($\beta=0.1$) in Figure \ref{2}(a), the slopes fitted to the numerical data points (dots) are $-0.93$ and $-1.96$. Recalling that in the non-interacting case, ${\rm Var}(a_2)\sim d^{-1}$ for initial random pure states and ${\rm Var}(a_2)\sim d^{-2}$ for random mixed states. Counterintuitively, for the low-temperature ($\beta=100$) case in Figure \ref{2}(b), the fitted slopes are approximately $-1.28$ and $-1.31$, which may suggest an even sharper concentration in this regime.

\textit{Universality of the typicality for random mixed states---}Beyond specific examples, we here provide general arguments, using insight from random matrix theory. For a large class of non-Hermitian random matrix such as complex Ginibre ensemble and random matrix whose entries are independent and identically distributed, the typical scaling of $O_k$ for \textit{bulk} eigenvalues and large $d$ was rigorously established as $O_k \sim \mathcal{O}(d)$ \cite{98prl_overlap,24jfm_bulk_overlap}. For random Lindbladian, this scaling has also been confirmed \cite{hamazaki22lindbladian,25randomLindblad}. This cannot guarantee the universality of the typicality for pure initial states.

However, if the initial state is a random mixed state, we expect that at least the concentration of the dominant relaxation mode ($a_2$) would be a universal behavior. The reason is that $\lambda_2$ should be classified as the \textit{edge} eigenvalue. Indeed, for many random matrix models, the typical scaling of $O_k$ of an edge eigenvalue $\lambda_k$ was shown to be $\mathcal{O}(\sqrt{d})$ \cite{15jmp_edge_overlap,23_edge_overlap}. Although for random Lindbladian, currently there is no rigorous proof, a typical scaling slower than $d$ can still be anticipated, which at least assures the concentration of $a_2$ for random mixed states. It is known that edge eigenvalues have larger eigenvalue spacing, which makes them more stable (smaller condition number). In a recent paper \cite{24eigenvector_scaling}, it was shown that a larger eigenvalue spacing implies a smaller dimensional scaling of the condition number $O_k$. 

\textit{A rigorous definition of the ``typical relaxation (mixing) time''---}The typical mixing time, $t^{\sigma_{\delta}}_{\epsilon}$, can be rigorously defined by first specifying a set of typical initial states, $\sigma_{\delta}$. It contains all initial states $\rho_0$ for which all $a_k$ (that are known to concentrate) are close to their steady-state expectation values: $\sigma_{\delta} := \{\rho_0 : |a_k(\rho_0) - \langle a_k \rangle| \leq \delta, \ \forall k \text{ s.t. } a_k \text{ concentrates}\}$. This definition excludes rare, atypical initial states that may exhibit anomalously large values of $|a_k|$. The typical relaxation time is then defined as: 
\begin{equation}\label{typical_mixing_time}
    t^{\sigma_{\delta}}_{\epsilon} := \inf \{t \ge 0 : \sup_{\rho_0 \in \sigma_{\delta}} \|e^{t\mathcal{L}}(\rho_0) - \rho_{\rm ss}\|_1 \leq \epsilon \}.
\end{equation}
This contrasts with the standard $\epsilon$-mixing time defined as
\begin{equation}
    t_{\epsilon}:= \inf \{t \ge 0 : \|e^{t\mathcal{L}}(\rho_0) - \rho_{\rm ss}\|_1 \leq \epsilon ,\ \forall \rho_0\}
\end{equation}
which considers the worst-case scenario over all possible initial states. Our definition recovers the standard one in the limit $\delta \to \infty$, where the set $\sigma_{\delta}$ encompasses all states. Consequently, the typical relaxation time is, by construction, a lower bound on the standard mixing time: $t_{\epsilon} = \lim_{\delta \to \infty} t^{\sigma_{\delta}}_{\epsilon} \ge t^{\sigma_{\delta}}_{\epsilon}$. This new definition may be useful for proving some weaker forms (but still physically relevant) of current results, such as the rapid mixing, under broader conditions. 

To further explore the implications of typicality for mixing times, let us define the $\epsilon$-mixing time for a single initial state $\rho$ as
\begin{equation}
t_{\epsilon}^{\rho}:= \inf \{t \ge 0 : \|e^{t\mathcal{L}}(\rho) - \rho_{\rm ss}\|_1 \leq \epsilon\}.
\end{equation}
A natural follow-up question is then to determine the conditions under which $t_{\epsilon}^{\rho}$ becomes independent of the initial state $\rho$ within the typical set $\sigma_{\delta}$, i.e., the mixing time itself exhibits initial-state typicality. Rigorous justification for cases like high-temperature thermalization, where all relaxation modes concentrate, is particularly relevant. Although a full proof is beyond the scope of this work, our results strongly suggest this is true, consistent with the observation that the full relaxation dynamics becomes irrelevant to the initial state in such cases (a relevant manuscript is in preparation). 

Additionally, note that when all $a_k$ concentrate around their means, $\langle a_k\rangle={\rm tr}(L_k^{\dagger})/d$, if $\delta$ is taken as an appropriate function of $d$ that vanishes as $d\to\infty$, the typical mixing time \eqref{typical_mixing_time} may be approximated by the $\epsilon$-mixing time for the maximally mixed state,
$t^{\mathbb{I}/d}_{\epsilon}$. This specific mixing time may be more amenable to analysis \cite{25arxiv_lin_rapid_ground}.

\textit{Re-evaluation of the efficiency of quantum Gibbs samplers---}The mixing time is essential in evaluating the efficiency of a quantum Gibbs sampler for efficient state preparations and quantum simulations. In addition to the conventional (worst-case) metric, it is natural to also report the \emph{typical} relaxation benchmark—namely the typical relaxation time defined above. With initial-state typicality, establishing efficiency may amount to showing \emph{typical rapid mixing}, i.e., that the typical relaxation time scales favorably, which is a weaker and more representative requirement than conventional rapid mixing. Even if a sampler is initialized atypically, a light randomization can push it into the typical set, as discussed in the main text. 

\end{document}


\title{Supplemental Material for ``Initial-State Typicality in Quantum Relaxation''
}

\author{Ruicheng Bao}
\email{Contact author: ruicheng@g.ecc.u-tokyo.ac.jp}
\affiliation{Department of Physics, Graduate School of Science, 
The University of Tokyo, Hongo, Bunkyo-ku, Tokyo 113-0033, Japan}

\maketitle

In this Supplemental Material, we provide some complementary derivations for the results presented in the main text. We also present some generalized results on initial-state typicality, including the typicality with a constraint on the initial state, and the approximate concentration under the approximate unitary 2-design.

\tableofcontents

\newpage

\vspace{12pt}

\section{Derivations of the first two moments of $a_k$}
\paragraph{Random pure state as the initial state---}Recall the definition is $a_k={\rm tr}(L_k^{\dagger}\rho_0)$, with $\rho_0=U\rho'_0U^{\dagger},\ U\sim {\rm Haar}(d)$. In what follows, we use $\langle \dots \rangle$ to denote the average over Haar measure. The average value of $a_k$ is 
\begin{equation}
    \langle a_k\rangle={\rm tr}(L_k^{\dagger}\langle\rho_0\rangle)={\rm tr}(L_k^{\dagger}\frac{\mathbb{I}}{d})=\frac{{\rm tr}(L_k^{\dagger})}{d}.
\end{equation}
We then calculate the variance of ${\rm tr}(L_k^{\dagger}\rho_0)$. First, we note that
\begin{equation}
    |a_k|^2={\rm tr}(L_k^{\dagger}\rho_0){\rm tr}(\rho_0 L_k)={\rm tr}(L_k^{\dagger}\rho_0){\rm tr}(L_k\rho_0)={\rm tr}[(L_k^{\dagger}\otimes L_k)(\rho_0 \otimes \rho_0)],
\end{equation}
where we use the relations ${\rm tr}(X){\rm tr}(Y)={\rm tr}(X\otimes Y)$ and $(AB)\otimes (CD)\equiv (A\otimes C)(B\otimes D)$. The second central moment is then given by 
\begin{equation}
    \langle  |a_k|^2\rangle={\rm tr}[(L_k^{\dagger}\otimes L_k)(\langle\rho_0 \otimes \rho_0\rangle)].
\end{equation}
One thus only needs to evaluate the quantity
\begin{equation}
    \langle\rho_0 \otimes \rho_0\rangle= \mathbb{E}_U[U^{\otimes 2}(\rho_0')^{\otimes 2}U^{\dagger,\otimes 2}].
\end{equation}
It is known that \cite{Mele2024introductiontohaar} for a Hermitian operator $O$,
\begin{equation}\label{second_moment}
    \mathbb{E}_U[U^{\otimes 2}OU^{\dagger,\otimes 2}]=c_{I,O}\mathbb{I}^{\otimes2}+c_{F,O}\mathbb{F},
\end{equation}
where $\mathbb{I}$ is the identity operator in the original Hilbert space, $\mathbb{F}\equiv \sum_{i,j}^d|i,j\rangle\langle j,i|$ is the swap (flip) operator and the coefficients are given by 
\begin{equation}
    c_{I,O}=\frac{{\rm tr}(O)-d^{-1}{\rm tr}(\mathbb{F}O)}{d^2-1}, \quad c_{F,O}=\frac{{\rm tr}(\mathbb{F}O)-d^{-1}{\rm tr}(O)}{d^2-1}.
\end{equation}
Substituting $O=(\rho'_0)^{\otimes 2}$ into the above equation yields
\begin{equation}\label{rho_square}
    \langle\rho_0 \otimes \rho_0\rangle=\frac{1-d^{-1}{\rm tr}(\rho_0'^2)}{d^2-1}\mathbb{I}^{\otimes2}+\frac{{\rm tr}(\rho_0'^2)-d^{-1}}{d^2-1}\mathbb{F}
\end{equation}
where ${\rm tr}[(\rho_0')^{\otimes 2}]=[{\rm tr}(\rho'_0)]^2=1$ and ${\rm tr}(\mathbb{F}(\rho_0')^{\otimes 2})={\rm tr}(\rho_0'^2)\leq 1$ have been used. The second equality is the outcome of the swap trick ${\rm tr}(XY)={\rm tr}[\mathbb{F}(X\otimes Y)]$. Consequently, we obtain
\begin{equation}
    \langle  |a_k|^2\rangle={\rm tr}[(L_k^{\dagger}\otimes L_k)(c_{I,\rho^{\otimes 2}}\mathbb{I}^{\otimes2}+c_{F,\rho^{\otimes 2}}\mathbb{F})]=c_{I,\rho^{\otimes 2}}|{\rm tr}(L_k^{\dagger})|^2+c_{F,\rho^{\otimes 2}}\|L_k^{\dagger}\|_{2}^2.
\end{equation}
Plugging this into the definition of variance, we get the desired result in the main text:
\begin{align}
    {\rm Var}_U(a_k)\equiv \langle  |a_k|^2\rangle -|\langle a_k\rangle|^2&=\left(c_{I,\rho^{\otimes 2}}-\frac{1}{d^2}\right)|{\rm tr}(L_k^{\dagger})|^2+c_{F,\rho^{\otimes 2}}\|L_k^{\dagger}\|_{2}^2\\
    &= \frac{[d^2{\rm tr}(\rho_0'^2)-d]\|L_k^{\dagger}\|_{2}^2+[1-d{\rm tr}(\rho_0'^2)]|{\rm tr}(L_k^{\dagger})|^2}{d^2(d^2-1)}\\
    &=\frac{[{\rm tr}(\rho_0'^2)-1/d][\|L_k^{\dagger}\|_{2}^2-|{\rm tr}(L_k^{\dagger})|^2/d]}{d^2-1}\geq0.
\end{align}
\paragraph{Random mixed state as the initial state.---}We next consider behaviors of $a_k$ in the ensemble of random mixed states, the Hilbert–Schmidt (HS) ensemble. 

The HS measure is induced by the partial trace operation. Assuming that the system of interest lives in the Hilbert space $\mathcal{H}_A$ with dimension $d$, we introduce an auxiliary Hilbert space $\mathcal{H}_B$ with dimension $d$. To realize random mixed state in the system of interest, we first prepare a random pure state $|\psi\rangle\sim {\rm Haar} (d^2)$ (extracted from the Haar measure) in the composite system $\mathcal{H}_A\otimes\mathcal{H}_B$. Then the random mixed state is given by 
\begin{equation}
    \rho_{0}\equiv {\rm tr}_B(|\psi\rangle\langle\psi|).
\end{equation}
We then compute the first two moments of $a_k\equiv{\rm tr}(L_k^{\dagger}\rho_{2})$, where the initial state $\rho_0=\rho_{2}$ is randomly taken from the HS measure. Because 
\begin{equation}
    \mathbb{E}[\rho_{0}]={\rm tr}_B\frac{\mathbb{I}_A\otimes \mathbb{I}_B}{d^2}=\frac{\mathbb{I}_A}{d},
\end{equation}
the first moment $\mathbb{E} (a_k)={\rm tr}(L_k^{\dagger})/d$ is the same as the Haar measure case. Following the same logic as the random pure state case, we obtain
\begin{equation}
    \mathbb{E}|a_k|^2={\rm tr}\left[(L_k^{\dagger}\otimes L_k)\mathbb{E}(\rho_{0}^{\otimes 2})\right]={\rm tr}\left[(L_k^{\dagger}\otimes L_k){\rm tr}_B\mathbb{E}_{|\psi\rangle}(|\psi\rangle\langle \psi|^{\otimes 2})\right].
\end{equation}
Using Eq. \eqref{rho_square}, we get
\begin{equation}
    \mathbb{E}_{|\psi\rangle}(|\psi\rangle\langle \psi|^{\otimes 2})=\frac{\mathbb{I}_{AB}^{\otimes2}+\mathbb{F}_{AB}}{d^2(d^2+1)},
\end{equation}
where we choose $\rho_0=|\psi\rangle\langle\psi|$ as a pure state by definition and notice the Hilbert space dimension is $d^2$ (dimension of the composite system). Notably, $\mathbb{I}_{AB}^{\otimes2}=\mathbb{I}_A\otimes \mathbb{I}_B\otimes\mathbb{I}_A\otimes \mathbb{I}_B$ and $\mathbb{F}_{AB}=\mathbb{F}_A\otimes \mathbb{F}_B$. By definitions of the identity operator and the flip operator, the partial trace over system $B$ yields:
\begin{equation}
    {\rm tr}_B\mathbb{E}_{|\psi\rangle}(|\psi\rangle\langle \psi|^{\otimes 2})=\frac{d^2 \mathbb{I}_A^{\otimes 2}+d\mathbb{F}_A}{d^2(d^2+1)}=\frac{d \mathbb{I}_A^{\otimes 2}+\mathbb{F}_A}{d(d^2+1)}.
\end{equation}
Consequently, 
\begin{align}
    &\mathbb{E}|a_k|^2=\frac{d|{\rm tr}(L_k^{\dagger})|^2+\|L_k\|_{2}^2}{d(d^2+1)}, \\
    &{\rm Var_{2}}(a_k)=\mathbb{E}|a_k|^2-|\mathbb{E}(a_k)|^2=\frac{\|L_k\|_{2}^2-|{\rm tr}(L_k^{\dagger})|^2/d}{d(d^2+1)}\lesssim d^{-3}O_k^{2}.
\end{align}
Notably, if the auxiliary space $\mathcal{H}_B$ has a different dimension $d_B\neq d$, the resulting measure ${\rm Ind}(d,d_B)$ induced by the partial trace is not the HS measure. If we randomly extract a mixed state $\rho_0$ from the induced measure ${\rm Ind}(d,d_B)$, the corresponding variance of $a_k(\rho_0)$ would be modified to 
\begin{equation}
    {\rm Var}_{\rho_0\sim {\rm Ind}(d,d_B)}[a_k(\rho_0)]=\frac{\|L_k\|_{2}^2-|{\rm tr}(L_k^{\dagger})|^2/d}{d(d\cdot d_B+1)}\lesssim \frac{O_k^2}{d^2d_B}.
\end{equation}

We now generalize the above results considered in the main text to a more general setting, where an arbitrary physical constraint is imposed on the initial state of the system of interest. The consideration is similar to the setting in \cite{06nphy_concentration}.

\section{Initial-state typicality under a general constraint: variance and bounds}

We now generalize the discussion to the setting where a global physical constraint selects a subspace $\mathcal H_R\subseteq\mathcal H_S\otimes\mathcal H_E$ with orthogonal projector $P_R$ and dimension $d_R={\rm tr}\,P_R$. A pure state is sampled uniformly (Haar) on $\mathcal H_R$:
\[
\rho:=|\phi\rangle\langle\phi|\,,\qquad |\phi\rangle\in\mathcal H_R,
\]
and the reduced state is $\rho_S={\rm tr}_E\rho$. For $L_k^\dagger$ acting on $\mathcal H_S$, define
\[
a_k:={\rm tr}_S(L_k^\dagger\,\rho_S)={\rm tr}_{SE}\!\big[(L_k^\dagger\!\otimes\!\mathbb I_E)\rho\big].
\]
Introduce the canonical objects associated with the constraint $R$:
\begin{equation}
\Omega_S:= {\rm tr}_E\!\Big(\frac{P_R}{d_R}\Big),\qquad
\Phi_R(X):=\frac{1}{d_R}\,{\rm tr}_E\!\big(P_R(X\!\otimes\!\mathbb I_E)P_R\big).
\label{eqS:defOmegaPhi}
\end{equation}
Here $\Omega_S$ is the average reduced state and $\Phi_R$ encodes the second-moment information.

\paragraph*{Second moment on the restricted subspace---}The derivations here are closely related to a representation theory formula used in \cite{06nphy_concentration}.
Let
\[
V:=\mathbb E_{\phi\in\mathcal H_R}\big[\,\rho^{\otimes 2}\,\big]
=\int_{\mathcal H_R} |\phi\phi\rangle\!\langle\phi\phi|\, d\phi
\ \in \ \mathsf{End}(\mathcal H_R\otimes\mathcal H_R).
\]
$V$ is invariant under the collective action $(U\otimes U)V(U^\dagger\otimes U^\dagger)=V$ for all $U\in{\rm U}(\mathcal H_R)$. By Schur--Weyl duality with $2$ factors, the commutant is spanned by the projectors onto the symmetric/antisymmetric subspaces of $\mathcal H_R\otimes\mathcal H_R$:
\[
\Pi^{(R)}_{\rm sym}=\tfrac12(\mathbb I_{RR'}+F_R),\qquad
\Pi^{(R)}_{\rm anti}=\tfrac12(\mathbb I_{RR'}-F_R),
\]
where $\mathbb I_{RR'}:=P_R\otimes P_R$ and 
\[
F_R:=(P_R\otimes P_R)\,F\,(P_R\otimes P_R)
\]
is the full-space swap $F:=\mathbb{F}_{SS'}\otimes\mathbb{F}_{EE'}$ compressed to $R\otimes R$. Since $\Pi^{(R)}_{\rm anti}|\phi\phi\rangle=0$ for all $|\phi\rangle$, the antisymmetric block does not contribute. Using ${\rm tr}\,V=1$ and $\dim{\rm Sym}^2(\mathcal H_R)=d_R(d_R+1)/2$, one obtains the second moment
\begin{equation}
\mathbb E\big[\rho^{\otimes 2}\big]
=\frac{2\,\Pi^{(R)}_{\rm sym}}{d_R(d_R+1)}
=\frac{(P_R\otimes P_R)+F_R}{d_R(d_R+1)}\,.
\label{eqS:secondmomentR}
\end{equation}

\paragraph*{Restricted-swap identity---}
For any $A,B$ acting on $\mathcal H_R$,
\begin{equation}
{\rm tr}\!\big[F_R(A\!\otimes\!B)\big]={\rm tr}\!\big[P_R\,A\,P_R\,B\big].
\label{eqS:restrictedswap}
\end{equation}
\emph{Proof.} Using $P_R^2=P_R$, $[F,P_R\otimes P_R]=0$, cyclicity of trace, and ${\rm tr}[(X\!\otimes\!Y)F]={\rm tr}(XY)$:
\[
\begin{aligned}
{\rm tr}\!\big[F_R(A\!\otimes\!B)\big]
&={\rm tr}\!\big[(P_R\!\otimes\!P_R)F(P_R\!\otimes\!P_R)(A\!\otimes\!B)\big]\\
&={\rm tr}\!\big[F(P_R A\!\otimes\!P_R B)\big]
={\rm tr}\!\big[(P_R A)(P_R B)\big]
={\rm tr}(P_R A P_R B).
\qquad\Box
\end{aligned}
\]

From \eqref{eqS:defOmegaPhi},
\[
\mathbb E[a_k]={\rm tr}_S(L_k^\dagger\,\Omega_S).
\]
Using \eqref{eqS:secondmomentR} and \eqref{eqS:restrictedswap} with $A=L_k^\dagger\!\otimes\!\mathbb I$, $B=L_k\!\otimes\!\mathbb I$,
\[
\begin{aligned}
\mathbb E|a_k|^2
&=\frac{1}{d_R(d_R+1)}
\Big(
{\rm tr}\big[(P_R\!\otimes\!P_R)(L_k^\dagger\!\otimes\!\mathbb I\otimes L_k\!\otimes\!\mathbb I)\big]
+{\rm tr}\big[F_R(L_k^\dagger\!\otimes\!\mathbb I\otimes L_k\!\otimes\!\mathbb I)\big]
\Big)\\
&=\frac{1}{d_R(d_R+1)}\Big(d_R^2\,|{\rm tr}(L_k^\dagger\Omega_S)|^2
+d_R\,{\rm tr}\big(L_k^\dagger\,\Phi_R(L_k)\big)\Big).
\end{aligned}
\]
Therefore
\begin{equation}
\boxed{\quad
{\rm Var}(a_k)
=\frac{{\rm tr}\!\big(L_k^\dagger\,\Phi_R(L_k)\big)-\big|{\rm tr}(L_k^\dagger\Omega_S)\big|^2}{d_R+1}\,.
\quad}
\label{eqS:var-exact-R}
\end{equation}
The numerator is nonnegative, since
\[
{\rm tr}\!\big(L_k^\dagger\,\Phi_R(L_k)\big)-|{\rm tr}(L_k^\dagger\Omega_S)|^2
=\frac{1}{d_R}\big\|\,P_R(L_k^\dagger\!\otimes\!\mathbb I)-{\rm tr}(L_k^\dagger\Omega_S)\,P_R\,\big\|_2^2\ge 0.
\]

\paragraph*{General upper bounds---}
Let $M:={\rm tr}_E P_R\ge 0$ on $\mathcal H_S$ with ${\rm tr}M=d_R$. Cauchy--Schwarz and $0\le M\le \|M\|_{\infty}\mathbb I$ give
\begin{equation}
{\rm tr}\!\big(L_k^\dagger\,\Phi_R(L_k)\big)
=\frac{1}{d_R}{\rm tr}_{SE}\!\big[(L_k^\dagger\!\otimes\!\mathbb I)P_R(L_k\!\otimes\!\mathbb I)P_R\big]
\le \frac{\|M\|_{\infty}}{d_R}\,\|L_k\|_2^2.
\label{eqS:keybound}
\end{equation}
Using \eqref{eqS:keybound} together with $|{\rm tr}(L_k^\dagger\Omega_S)|^2\ge 0$ in \eqref{eqS:var-exact-R} yields
\begin{equation}
{\rm Var}(a_k)\ \le\ \frac{\kappa_R}{d_R+1}\,\|L_k\|_2^2,\qquad
\kappa_R:=\frac{\|\,{\rm tr}_E P_R\,\|_{\infty}}{d_R}\in\Big[\frac{1}{d_S},\,1\Big].
\label{eqS:HSbound}
\end{equation}
A cruder but dimension-agnostic bound follows from ${\rm tr}(L_k^\dagger\Phi_R(L_k))\le \|L_k\|_\infty^2$:
\begin{equation}
{\rm Var}(a_k)\ \le\ \frac{\|L_k\|_\infty^2}{d_R+1}\leq \frac{\|L_k\|_2^{2}}{d_R+1}\leq\frac{O_k^2d_S}{d_R+1}.
\label{eqS:opbound}
\end{equation}
\emph{Consistency checks.}
(i) If $P_R=\mathbb I_{SE}$ (no constraint; $d_R=d_S d_E$), then
$\Omega_S=\mathbb I_S/d_S$, $\Phi_R(X)=X/d_S$ and \eqref{eqS:var-exact-R} reduces to
\begin{equation}
{\rm Var}(a_k)
=\frac{\|L_k\|_2^2-\dfrac{|{\rm tr}L_k^\dagger|^2}{d_S}}
{d_S(d_S d_E+1)}.
\label{eqS:induced}
\end{equation}
(ii) If $P_R=\mathbb I_S\otimes P_E$ (constraint only on environment, isotropic on $S$), then
$M=\tfrac{d_R}{d_S}\mathbb I_S$, hence $\kappa_R=\tfrac{1}{d_S}$ and
\[
{\rm Var}(a_k)\le \frac{\|L_k\|_2^2}{d_S(d_R+1)}.
\]
(iii) If $P_R=P_S\otimes |\mathrm e_0\rangle\langle \mathrm e_0|$,
then $M=P_S$ and $\kappa_R=\tfrac{1}{d_R}$, giving
\[
{\rm Var}(a_k)\le \frac{\|L_k\|_2^2}{d_R(d_R+1)}\sim \frac{1}{d_R^2}.
\]

From the expression of the upper bound of ${\rm Var}(a_k)$, Eq. \eqref{eqS:opbound}, it is clear that the initial-state typicality requires $d_R\gg d_S$. This is usually the case for a conventionally understanding ``environment'' coupled to the system of interest.

\paragraph*{Special case $d_E=1$ (constraint only on the system).}
Here $\mathcal H_R\subseteq\mathcal H_S$, $P_R$ projects on $\mathcal H_S$ and $d_R={\rm tr}\,P_R$. Then
\[
\Omega_S=\frac{P_R}{d_R},\qquad
\Phi_R(X)=\frac{1}{d_R}\,P_R X P_R.
\]
Let $(L_k^\dagger)_R:=P_R L_k^\dagger P_R$. The exact variance and a useful upper bound are
\begin{equation}
{\rm Var}(a_k)
=\frac{{\rm tr}\!\big((L_k^\dagger)_R (L_k^\dagger)_R^{\dagger}\big)-\dfrac{|{\rm tr}(L_k^\dagger)_R|^2}{d_R}}
{d_R(d_R+1)},
\qquad
{\rm Var}(a_k)\le
\frac{\|L_k\|_2^2}{d_R(d_R+1)}.
\label{eqS:dE1}
\end{equation}
The HS bound is tight whenever ${\rm tr}(L_k^\dagger)_R=0$. Therefore, even if the constraint is solely on the system of interest, the initial-state typicality can still occur given that $d_R$ is close to the system's Hilbert space dimension $d$.

\paragraph*{Two identities used repeatedly.}
For any $X$ on $\mathcal H_S$ and $M={\rm tr}_E P_R$,
\[
\|(X\!\otimes\!\mathbb I)P_R\|_2^2
={\rm tr}\big(P_R(X^\dagger X\!\otimes\!\mathbb I)P_R\big)
\le {\rm tr}(X^\dagger X\,M)\le \|M\|_{\infty}\,\|X\|_2^2,
\qquad
{\rm tr}(L_k^\dagger\Omega_S)=\frac{1}{d_R}\,{\rm tr}_{SE}\big[(L_k^\dagger\!\otimes\!\mathbb I)P_R\big].
\]

\section{Details of the non-interacting dissipative qubits system considered in Example 1 of the main text}
For an individual dissipative qubit, the Hamiltonian and jump operators given by:
\begin{align}
 H &= \frac{E}{2} \sigma_z, \\
 J_1 &= \sqrt{\gamma_1} \sigma_-, \quad J_2 = \sqrt{\gamma_0} \sigma_+.
\end{align}
Here, $J_1$ describes the energy decay from the excited state to the ground state at a rate $\gamma_1$, while $J_2$ represents an incoherent pumping process from the ground state to the excited state at a rate $\gamma_0$. For this model, the matrix representation of a single-qubit Lindbladian is given by
\begin{equation}
\mathcal{L}=\left(\begin{array}{cccc}
-\gamma_{0} & 0 & 0 & \gamma_{1}\\
0 & -iE-\frac{\gamma_{0}+\gamma_{1}}{2} & 0 & 0\\
0 & 0 & iE-\frac{\gamma_{0}+\gamma_{1}}{2} & -0\\
\gamma_{0} & 0 & 0 & -\gamma_{1}
\end{array}\right).
\end{equation}

The eigenmatrices are uniquely determined by the biorthogonality condition $\text{tr}[L_k^{\dagger}R_h]=\delta_{kh}$ and the normalization constraints mentioned in the text. The eigenvalues, along with the corresponding right and left eigenmatrices of the Lindbladian, are given by.
\begin{align}
\lambda_{1} & = 0, & R_{1} & = \frac{1}{\gamma_{0}+\gamma_{1}}\begin{pmatrix} \gamma_{1} & 0\\ 0 & \gamma_{0} \end{pmatrix}, & L_{1} & = \begin{pmatrix} 1 & 0\\ 0 & 1 \end{pmatrix} \\
\lambda_{2} & = -\frac{\gamma_{0}+\gamma_{1}}{2}-iE, & R_{2} & = \begin{pmatrix} 0 & 1\\ 0 & 0 \end{pmatrix}, & L_{2} & = \begin{pmatrix} 0 & 0\\ 1 & 0 \end{pmatrix} \\
\lambda_{3} & = -\frac{\gamma_{0}+\gamma_{1}}{2}+iE, & R_{3} & = \begin{pmatrix} 0 & 0\\ 1 & 0 \end{pmatrix}, & L_{3} & = \begin{pmatrix} 0 & 1\\ 0 & 0 \end{pmatrix} \\
\lambda_{4} & = -(\gamma_{0}+\gamma_{1}), & R_{4} & = \frac{1}{2}\begin{pmatrix} 1 & 0\\ 0 & -1 \end{pmatrix}, & L_{4} & = \frac{2}{\gamma_{0}+\gamma_{1}}\begin{pmatrix} \gamma_{1} & 0\\ 0 & -\gamma_{0} \end{pmatrix}
\end{align}
As a result, we obtain the HS norms of these left eigenmatrices as 
\begin{equation}
    \|L_1\|_2 = \sqrt{2}, \quad \|L_2\|_2 = 1, \quad \|L_3\|_2 = 1, \quad \|L_4\|_2 = \frac{2\sqrt{\gamma_0^2 + \gamma_1^2}}{\gamma_0+\gamma_1}.
\end{equation}

For a system of $N$ non-interacting qubits, the total Lindbladian $\mathcal{L}^{\rm tot}$ is the sum of the Lindbladians for each individual qubit, $\mathcal{L} = \sum_{i=1}^N \mathcal{L}^{(i)}$, where $\mathcal{L}^{(i)} = \mathbb{I}_1 \otimes \dots \otimes \mathcal{L}_{i} \otimes \dots \otimes \mathbb{I}_N$ acts only on the $i$-th qubit's Hilbert space. Let $R_{\vec{k}} = \bigotimes_{j=1}^N R_{k_j}$ be a tensor-product eigenmatrix of the N-qubit system, where for a single qubit $\mathcal{L}_{\text{single}}(R_{k_j}) = \lambda_{k_j} R_{k_j}$. Applying the total Lindbladian to $R_{\vec{k}}$ yields:

\begin{align}
\mathcal{L}^{\rm tot}(R_{\vec{k}}) &= \left(\sum_{i=1}^N \mathcal{L}^{(i)}\right) \left(\bigotimes_{j=1}^N R_{k_j}\right) 
= \sum_{i=1}^N \lambda_{k_i} \left(\bigotimes_{j=1}^N R_{k_j}\right) = \left(\sum_{i=1}^N \lambda_{k_i}\right) R_{\vec{k}}. 
\end{align}
This explicitly shows that the eigenvalues $\Lambda_{\vec{k}}$ and the corresponding right ($R_{\vec{k}}$) and left ($L_{\vec{k}}$) eigenmatrices for $\mathcal{L}^{\rm tot}$ are given by the sum of single-qubit eigenvalues and the tensor product of single-qubit eigenmatrices:
\begin{equation}
    \Lambda_{\vec{k}} = \sum_{i=1}^{N} \lambda_{k_i}, \quad R_{\vec{k}} = \bigotimes_{i=1}^{N} R_{k_i}, \quad L_{\vec{k}} = \bigotimes_{i=1}^{N} L_{k_i},
\end{equation}
where we define $\vec{k}=(k_1,k_2,...,k_N)$ with $k_i\in\{1,2,3,4\}$.

The HS norm of a tensor-product operator is the product of the individual norms. Therefore, the norm of the N-qubit left eigenmatrix $L^{\rm tot}_{\vec{k}}$ is $ \|L^{\rm tot}_{\vec{k}}\|_2 = \prod_{i=1}^{N} \|L_{k_i}\|_2$. The slowest decaying modes correspond to one qubit relaxing in its slowest mode (eigenvalue $\lambda_2$ or $\lambda_3$) while the other $N-1$ qubits are in the steady state (eigenvalue $\lambda_1 = 0$). For these modes, for instance, with the multi-index $\vec{k}=(2,1,\ldots,1)$ and its permutations, the HS norm of the corresponding left eigenmatrix is:
\begin{equation}
    \|L^{\rm tot}_{(2,1,\ldots,1)}\|_2 = \|L_2\|_2 \cdot \prod_{i=2}^{N} \|L_1\|_2 = 1 \cdot (\sqrt{2})^{N-1} = 2^{(N-1)/2}\sim \sqrt{d}.
\end{equation}

Next, we discuss the properties of $\|L_4\|_2$ so that we can clearly see the behaviors of other relaxation modes. Its value lies in $[\sqrt{2}, 2]$. It reaches its minimum value of $\sqrt{2}$ when $\gamma_1 = \gamma_0$, and it approaches its supremum of $2$ when the ratio of the rates diverges, e.g., when one rate is zero while the other is not. We then introduce a more specific physical scenario. Assume that the rates $\gamma_1$ and $\gamma_0$ for each qubit are determined by a heat bath at an inverse temperature $\beta$ and satisfy the local detailed balance condition: $\gamma_0 / \gamma_1 = e^{-\beta E}$, where $E$ is the energy gap. In this context, the magnitude of $\|L_4\|_2$ is directly dependent on temperature. Only in the zero-temperature limit ($\beta \to \infty$), the rate ratio $\gamma_1/\gamma_0$ diverges, causing $\|L_4\|_2 \to 2$. In this limit, the maximum HS norm of any left eigenmatrix of the N-qubit system scales as $\max_{\vec{k}} \|L_{\vec{k}}\|_2 = (\|L_4\|_2)^N \to 2^N = d$, where $d=2^N$. For any non-zero temperature ($T>0, \beta < \infty$), the value of $\|L_4\|_2$ is strictly bounded between $\sqrt{2}$ (corresponding to the $T \to \infty$ limit) and $2$. Consequently, the scaling of the maximum norm is also strictly bounded between $(\sqrt{2})^N = d^{1/2}$ and $2^N = d$.

Therefore, for this interacting model, $a_k$ concentrates around their mean for all $k$ in the thermodynamic limit for any finite $\gamma_0,\ \gamma_1$. That is, the full relaxation dynamics becomes independent of the initial state when $N\to\infty$. 

Then, we analyze the behaviors of $a_{\vec k}\approx \langle a_{\vec k}\rangle$ (typical value) and $a^{\max}_{\vec k}:=\max_{\rho}|a_{\vec k}|$ (maximum value), where $a_{\vec k}={\rm tr}(L_{\vec k}^{\rm tot,\dagger}\rho)$ and $\rho$ is the initial state. 
We first consider the maximal behavior. Assuming $\vec k=(k_1,\ldots,k_N)$ has $m_\alpha$ sites of type $\alpha\in\{1,2,3,4\}$, then
the maximization over $\rho$ gives the numerical radius
\begin{equation}
a_{\vec k}^{\max}=\max_{\rho}|{\rm tr}(L_{\vec k}^{\rm tot,\dagger}\rho)|
=:w(L_{\vec k}^{\rm tot})
=\max_{\theta}\lambda_{\max}\!\left[\frac{e^{-i\theta}L_{\vec k}^{\rm tot}
+e^{i\theta}L_{\vec k}^{\rm tot,\dagger}}{2}\right].
\end{equation}
When $\vec k$ contains only $L_1$ and $L_4$ (i.e., $m_2+m_3=0$), 
$L_{\vec k}^{\rm tot}$ is diagonal Hermitian and 
$w(L_{\vec k}^{\rm tot})=c^{m_4}$ with 
$c=2\max(\gamma_0,\gamma_1)/(\gamma_0+\gamma_1)\in[1,2)$ being the maximum eigenvalue of $L_4$. 
If at least one $L_{2,3}$ appears, 
$w(L_{\vec k}^{\rm tot})=\tfrac12\,c^{m_4}$. 
Hence the global maximum over all non-stationary modes is
\begin{equation}
\max_{\vec k\neq(1,\ldots,1),\,\rho}|a_{\vec k}|
=\max_{\vec k}a_{\vec k}^{\max}
=c^{N}
=\left(\frac{2\max(\gamma_0,\gamma_1)}{\gamma_0+\gamma_1}\right)^{\!N}
=\mathcal{O}(d^{\log_2 c}),
\label{eq:a-max-global}
\end{equation}
which can reach $\mathcal{O}(d)$ for strongly unbalanced rates. Specifically, for the slowest mode $\vec k =(2,1,...,1)$, we have $m_4=0$, so that $a_{(2,1,...,1)}^{\max}=\frac{1}{2}.$ This is in contrast with the typical value, $a_{(2,...,1)}\approx \langle a_{(2,...,1)}\rangle=0$. Specifically, the scaling of the typical value is $\sqrt{\langle|a_{(2,...,1)}|^2\rangle}\sim \mathcal{O}(d^{-1/2})$ for Haar ensemble and $\mathcal{O}(d^{-1})$ for HS ensemble.

We next consider the typical value $\langle a_{\vec k}\rangle$ maximized over $\vec k$, which is evaluated by choosing $\rho$ as the maximally mixed state $\rho=\mathbb{I}/d$:
\begin{equation}
\langle a_{\vec k}\rangle =\frac{1}{d}\,{\rm tr}\!\left(L_{\vec k}^{\rm tot,\dagger}\right)
=\left(\frac{\tau}{2}\right)^{m_4},
\quad
\tau={\rm tr}(L_4)=\frac{2(\gamma_1-\gamma_0)}{\gamma_0+\gamma_1},
\end{equation}
and $a_{\vec k}=0$ whenever $m_2+m_3\ge1$. 
Excluding the steady-state, the largest value occurs at $m_4=1$ (because $|\frac{\tau }{2}|\leq 1$), giving
\begin{equation}
\max_{\vec k\neq(1,\ldots,1)}
\langle a_{\vec k}\rangle
=\frac{\gamma_1-\gamma_0}{\gamma_0+\gamma_1}
=\mathcal{O}(1).
\label{eq:a-mix-max}
\end{equation}

We thus prove the claim in the main text that $a_k\approx \langle a_k\rangle$ is at most of the order $\mathcal{O}(1)$, while $\max_{\rho_0}|a_k|$ can scale up to $\mathcal{O}(d)$. That is, there is a strong separation between the maximal relaxation time and the typical relaxation time. The maximal relaxation time may not capture the typical behavior of relaxation in many-body open quantum systems.

Finally, as mentioned in the main text, if we further add the dephasing noise to each qubit, the above discussion will not be affected. Adding the dephasing noise simply contributes an additional term $\mathcal{L}_z$ to the original single-qubit Lindbladian $\mathcal{L}$, where 
\begin{equation}
    \mathcal{L}_z(\rho)=\gamma_D(\sigma_z\rho \sigma_z-\rho).
\end{equation}
This additional term commutes with the original Lindbladian, $[\mathcal{L},\mathcal{L}_z]=0$, so they have the same eigenbasis. As a result, all eigenmatrices presented before are invariant. $\lambda_1$ and $\lambda_4$ are also invariant. The only difference is that $\lambda_{2,3}\to \lambda_{2,3}-2\gamma_D$. That is, the spectral gap will be determined by the original $\lambda_4=-(\gamma_0+\gamma_1)$, if the dephasing rate $\gamma_D$ is sufficiently large, i.e., $\gamma_D>(\gamma_0+\gamma_1)/2$. 

\section{Rapid Mixing Implies Concentration of Relaxation Modes}

Here, we provide a self-contained argument demonstrating that rapid mixing is a sufficient condition for the concentration of relaxation modes under a physical assumption. 

To formalize rapid mixing, one first defines the $\epsilon$-mixing time $t_{\epsilon}$ based on the trace norm, which tracks the worst-case distance from the steady state:
\begin{equation}
    t_{\epsilon}:=\inf_{t}\{t\geq 0:\|e^{s\mathcal{L}}(\rho_0)-\rho_{\rm ss}\|_1\leq \epsilon,\ \forall\rho_0,\ s\geq t \}.
\end{equation}
The rapid mixing condition states that for any $\epsilon>0$, $t_{\epsilon}$ scales only polylogarithmically with dimension $d$:
\begin{equation}
    t_{\epsilon}\sim\mathcal{O}(\log({\rm poly} [\log d]/\epsilon)).
\end{equation}
This condition can be equivalently stated as \cite{cubitt15rapidmixing}: there exist positive and size-independent constants $g$, $\alpha$, $c$ such that
\begin{equation}
    \|e^{t\mathcal{L}}(\rho_0)-\rho_{\rm ss}\|_1\leq c[\log(d)]^{\alpha}e^{-gt}
\end{equation}
holds for any $\rho_0$ and $t\geq0$.

To connect this macroscopic definition to the microscopic overlaps $a_k$, we must relate $t_{\epsilon}$ to the worst-case timescale derived from the spectral decomposition. Let us define this ``spectral relaxation time'' $t_{\rm spec}(\epsilon)$ as:
\begin{equation}
    t_{\rm spec}(\epsilon):=\max_k \sup_{\rho_0}\left\{\frac{1}{|{\rm Re}(\lambda_k)|}\log{\left(\frac{|a_k(\rho_0)|}{\epsilon}\right)}\right\}.
\end{equation}
This is determined from the estimation $|a_k|e^{-|{\rm Re}(\lambda_k)|t_{\epsilon}}\leq \epsilon, \forall k$ adopted in \cite{Mori20resolving}. Our key assumption, which has been numerically verified in \cite{Mori20resolving}, is that these two timescales scale identically with the system dimension $d$: $t_{\epsilon} \sim t_{\rm spec}(\epsilon)$.

This assumption is physically well-motivated. From the spectral decomposition, $\|e^{s\mathcal{L}}(\rho_0)-\rho_{\rm ss}\|_1=\|\sum_{k=2}^{d^2}a_ke^{\lambda_kt}R_k \|_1$ sums over $\sim d^2$ relaxation modes. For this quantity to be bounded by a polylogarithmic term, the vast majority of these $d^2$ contributions must be small. Imagine, for contradiction, that a small but finite fraction $f > 0$ of the $d^2$ modes had large worst-case overlaps. The number of such modes would be $f d^2$. Even if each overlap is not individually large, their collective contribution to the trace norm would almost certainly scale with $d^2$, trivially violating the rapid mixing condition. Therefore, the fraction of modes with large overlaps cannot be a fixed constant; this fraction must itself vanish as $d \to \infty$. This implies that $\sup_{\rho_0}|a_k(\rho_0)|$ must be sufficiently bounded for almost all relevant modes.

Under this assumption, the rapid mixing condition (Eq. S2) implies $t_{\rm spec}(\epsilon) \sim \mathcal{O}(\log({\rm poly} [\log(d)]/\epsilon))$. This, in turn, requires that the worst-case overlap for any mode $k$ must be similarly bounded:
\begin{equation}
    \sup_{\rho_0} |a_k(\rho_0)| \lesssim \mathcal{O}({\rm poly} [\log d]).
\end{equation}
The second ingredient is the numerical radius inequality:
\begin{equation}
    \max_{\rho_0}|a_k|=
    \max_{|\phi\rangle}|\langle \phi|L_k^{\dagger}|\phi\rangle| \geq\frac{\|L_k^{\dagger}\|_{\rm op}}{2}\geq \frac{\|L_k^{\dagger}\|_{2}}{2\sqrt{d}},
\end{equation}
which implies 
\begin{equation}
    \|L_k\|_2\leq 2\sqrt{d}\max_{\rho_0}|a_k|.
\end{equation}
Combining these two results, we obtain a bound on the scaling of $\|L_k\|_2$:
\begin{equation}
    \|L_k\|_2 \lesssim \mathcal{O}(\sqrt{d}\ {\rm poly} [\log d]).
\end{equation}
Finally, we insert this bound into our main text formulas for the variance. For random pure states:
\begin{equation}
    {\rm Var}_U(a_k) < \frac{\|L_k\|_2^2}{d^2} \lesssim \frac{\mathcal{O}(d \cdot {\rm poly}[\log d])}{d^2} \sim \mathcal{O}\left(\frac{{\rm poly}[\log d ]}{d}\right).
\end{equation}
For random mixed states, the bound is even stronger:
\begin{equation}
    {\rm Var}(a_k) < \frac{\|L_k\|_2^2}{d^3} \lesssim \frac{\mathcal{O}(d \cdot {\rm poly}[\log(d)])}{d^3} \sim \mathcal{O}\left(\frac{{\rm poly}[\log(d)]}{d^2}\right).
\end{equation}
Since both variances vanish rapidly as $d\to\infty$, we conclude that rapid mixing is a sufficient condition for the concentration of relaxation modes.

\section{Derivation of the upper bound $O_k\leq \sqrt{p_{\rm max}/p_{\rm min}}$ for quantum detailed balance systems}

In this section, we provide a detailed derivation for the upper bound of the overlap between the left and right eigenvectors of a Liouvillian satisfying quantum detailed balance (QDB).
\subsection{Davies map and the GNS condition}
We consider a quantum system described by a Lindblad master equation. The Liouvillian super-operator $\mathcal{L}$ is defined as:
\begin{equation}
    \mathcal{L} = i\mathcal{U} + \mathcal{D},
\end{equation}
where the unitary part is generated by $\mathcal{U}(X) = -[H, X]$ with $H$ being the system Hamiltonian, and $\mathcal{D}$ is the dissipative part. The system is assumed to thermalize to a unique Gibbs state $\rho_\beta = e^{-\beta H} / {\rm tr}(e^{-\beta H})$ at inverse temperature $\beta$. The maximum and minimum eigenvalues of $\rho_\beta$ are denoted by $ p_{\max}$ and $ p_{\min}$, respectively.

The right and left eigenvectors of $\mathcal{L}$, denoted $R_k$ and $L_k$, are defined by $\mathcal{L}(R_k) = \lambda_k R_k$ and $\mathcal{L}^\dagger(L_k) = \lambda_k^* L_k$. The adjoint $\dagger$ is taken with respect to the HS inner product $\langle A, B \rangle_{2} = {\rm tr}(A^\dagger B)$. The eigenvectors are assumed to satisfy the biorthonormalization condition ${\rm tr}(L_j^\dagger R_k) = \delta_{jk}$.

The derivation relies on two physical assumptions for the Davies map:
\begin{enumerate}
    \item GNS QDB: The dissipative part $\mathcal{D}$ is self-adjoint with respect to the weighted inner product $\langle A, B \rangle_{\rho_\beta} = {\rm tr}(\rho_\beta A^\dagger B)$.
    \item Commutativity: The generator of unitary evolution $\mathcal{U}$ and the dissipator $\mathcal{D}$ commute, i.e., $[\mathcal{U}, \mathcal{D}] = 0$.
\end{enumerate}
To proceed, we need the following intermediate results.

{\textit{Lemma 1}---}The left and right eigenoperator of the Lindbladian $\mathcal{L}$ satisfies the following relation given that the corresponding eigenvalue $\lambda_k$ is non-degenerate (this constraint can be relaxed, see the end of this section):
\begin{equation}
    L_k = c_k R_k \rho_\beta,
    \label{eq:eigenvector_relation_L}
\end{equation}
where $c_k$ is a normalization constant.

{\textit{Proof}---}The GNS-QDB condition for $\mathcal{D}$ states that $\langle A, \mathcal{D}(B) \rangle_{\rho_\beta} = \langle \mathcal{D}(A), B \rangle_{\rho_\beta}$ for all operators $A, B$. This self-adjoint condition can be shown to be equivalent to the operator identity 
\begin{equation}
    \mathcal{D}^\dagger(X) = \mathcal{D}(X\rho_\beta^{-1})\rho_\beta.
\end{equation}
Let $L_k^\mathcal{D}$ be a left eigenvector of $\mathcal{D}$ with a real eigenvalue $\nu_k$. The derivation proceeds as follows:
\begin{align}
    \mathcal{D}^\dagger(L_k^\mathcal{D}) = \nu_k L_k^\mathcal{D}  
    \implies \mathcal{D}((L_k^\mathcal{D})\rho_\beta^{-1})\rho_\beta = \nu_k L_k^\mathcal{D}  
    \implies \mathcal{D}((L_k^\mathcal{D})\rho_\beta^{-1}) = \nu_k (L_k^\mathcal{D}\rho_\beta^{-1}).
\end{align}
The final line shows that the operator $(L_k^\mathcal{D})\rho_\beta^{-1}$ is a right eigenvector of $\mathcal{D}$ with the same eigenvalue $\nu_k$. Therefore, it must be proportional to the corresponding right eigenvector $R_k^\mathcal{D}$. This establishes the relation:
\begin{equation}
    L_k^\mathcal{D} \propto R_k^\mathcal{D}\rho_\beta.
    \label{eq:eigenvector_relation_D}
\end{equation}

Further, the commutativity condition $[\mathcal{U}, \mathcal{D}]=0$ guarantees that $\mathcal{U}$ and $\mathcal{D}$ share a common basis of eigenvectors. These eigenvectors are necessarily also the eigenvectors of the full Liouvillian $\mathcal{L} = i\mathcal{U} + \mathcal{D}$. Specifically, if we consider the case where the eigenvalues $\lambda_k$ of the full Liouvillian $\mathcal{L}$ are non-degenerate, the situation simplifies. Each eigenvector $R_k$ of $\mathcal{L}$ is unique (up to a scalar). Since $[\mathcal{L}, \mathcal{D}] = i[\mathcal{U}, \mathcal{D}] + [\mathcal{D}, \mathcal{D}] = 0$, any eigenvector of $\mathcal{L}$ must also be an eigenvector of $\mathcal{D}$. The same logic applies to the left eigenvectors. Therefore, the eigenvectors of $\mathcal{L}$ are the same as the shared eigenvectors of $\mathcal{U}$ and $\mathcal{D}$. We thus finish the proof of Lemma 1.

\textit{Lemma 2---}The H$\ddot{\rm o}$lder inequality for Schatten-$p$ norm:
\begin{equation}\label{Holder}
    \| AB\|_r\leq \|A\|_p\|B\|_q,
\end{equation}
for any operators $A,\ B\in \mathcal{H}$, where $p,q,r\in [1,\infty]$ and satisfy $\frac{1}{p}+\frac{1}{q}=\frac{1}{r}$.

With these Lemmas, we can derive the desired upper bound. We start with the relationship from Eq.~\eqref{eq:eigenvector_relation_L} and use the biorthonormalization condition ${\rm tr}(L_k^\dagger R_k)=1$ to find the constant $c_k$.
\begin{align}
    {\rm tr}[(c_k R_k \rho_\beta)^\dagger R_k] = 1 
    \implies c_k^* {\rm tr}(\rho_\beta R_k^\dagger R_k) = 1,
\end{align}
since $\rho_\beta$ is Hermitian. This gives $|c_k| = 1 / |{\rm tr}(\rho_\beta R_k^\dagger R_k)|$.

The product of the HS norms is then given by
\begin{equation}
    \|L_k\|_{2} \|R_k\|_{2} = |c_k| \|R_k \rho_\beta\|_{2} \|R_k\|_{2} 
    = \frac{\|R_k \rho_\beta\|_{2} \|R_k\|_{2}}{|{\rm tr}(\rho_\beta R_k^\dagger R_k)|}.
\end{equation}
We now bound the numerator from above and the denominator from below.

First, we derive a lower bound for the denominator. The lower bound is obtained by expanding the trace
\begin{align}
    |{\rm tr}(\rho_\beta R_k^\dagger R_k)| = {\rm tr}(\rho_\beta R_k^\dagger R_k) &=\sum_{i,j}\langle i|\rho_{\beta}R_k^{\dagger}|m\rangle\langle m|R_k|i\rangle\nonumber\\&=\sum_{i,j}p_i \langle i|R_k^{\dagger}|m\rangle\langle m|R_k|i\rangle
    \ge  p_{\min}(\rho_\beta){\rm tr}(R_k^\dagger R_k) =  p_{\min}\|R_k\|_{2}^2.
\end{align}
where we insert a $\sum_{j}|j\rangle\langle j|=\mathbb{I}$ and use $\rho_{\beta}|i\rangle=p_i|i\rangle$.
The next step is to upper bound the numerator. Using the H$\ddot{\rm o}$lder inequality in Lemma 2 \eqref{Holder}, we get 
\begin{align}
    \|R_k \rho_\beta\|_{2} = \|R_k \rho_\beta^{1/2}\rho_\beta^{1/2}\|_{2}\leq \|R_k \rho_\beta^{1/2}\|_2\|\rho_\beta^{1/2}\|_{\infty}=\sqrt{p_{\max}(\rho_\beta)}\|R_k \rho_\beta^{1/2}\|_2.
\end{align}
Noticing that by definition and the cyclic property of the trace,
\begin{equation}
    \|R_k \rho_\beta^{1/2}\|_2^2={\rm tr}(\rho_{\beta}^{1/2}R_k^{\dagger}R_k\rho_{\beta}^{1/2})={\rm tr}(\rho_{\beta}R_k^{\dagger}R_k),
\end{equation}
which is exactly the denominator. Thus, the upper bound of the numerator can be rewritten as $\sqrt{p_{\max}(\rho_{\beta}){\rm tr}(\rho_{\beta}R_k^{\dagger}R_k)}.$

Then, combining the bounds for the numerator and the denominator, we arrive at the final result:
\begin{align}
    O_k\equiv\|L_k\|_{2} \|R_k\|_{2} \le \frac{\sqrt{p_{\max}(\rho_{\beta}){\rm tr}(\rho_{\beta}R_k^{\dagger}R_k)}}{ {\rm tr}(\rho_{\beta}R_k^{\dagger}R_k)} \|R_k\|_2
    = \frac{ \sqrt{p_{\max}}}{ \sqrt{{\rm tr}(\rho_{\beta}R_k^{\dagger}R_k)}}\|R_k\|_2\leq\frac{\sqrt{p_{\max}}}{\sqrt{p_{\min}}}.
\end{align}

\subsection{Upper bound under the KMS condition}
The KMS inner product is defined as
\begin{equation}
    \langle A,B\rangle_{\rm KMS}:=\langle\rm A,\mathcal{G}(B)\rangle_2,\ \mathcal{G}(A):=\rho_{\beta}^{1/2}A\rho_{\beta}^{1/2}.
\end{equation} 
The KMS-QDB condition is the the Lindbladian $\mathcal{L}$ is self-adjoint with respect to the KMS inner product, $\langle A, \mathcal{L}(B) \rangle_{\rm KMS} = \langle \mathcal{L}(A), B \rangle_{\rm KMS}$ for all operators $A, B$. The KMS-QDB condition is provably broader than the GNS-QDB condition, that is, the class of Lindbladian with KMS-QDB condition is strictly larger than the class of Lindbladian satisfying GNS-QDB condition \cite{carlen2017gradient,lin_25cmp_rapidmixing}. The KMS-QDB condition is equivalent to the operator equality:
\begin{equation}
    \mathcal{L}=\mathcal{G}^{-1} \circ\mathcal{L}^{\dagger}\circ \mathcal{G}.
\end{equation}
For simplicity, we first consider the cases where $\lambda_k$ is non-degenerate. Acting $\mathcal{L}$ on its right eigenvector $R_k$ then yields
\begin{equation}
    \mathcal{L}(R_k)=\mathcal{G}^{-1} \circ\mathcal{L}^{\dagger}\circ \mathcal{G}(R_k)=\lambda_k R_k\Rightarrow \mathcal{L}^{\dagger}[\mathcal{G}(R_k)]=\lambda_k\mathcal{G}(R_k),
\end{equation}
which implies that $\mathcal{G}(R_k)$ is the $k$th eigenvector $L_k$ according to its definition. Imposing the biorthogonal condition leads to the expressions of $L_k$ and its HS norm:
\begin{equation}
    L_k=\frac{\rho_{\beta}^{1/2}R_k\rho_{\beta}^{1/2}}{{\rm tr}(\rho_{\beta}^{1/2}R^{\dagger}_k\rho_{\beta}^{1/2}R_k)},\quad \|L_k\|_2=\frac{\|\rho_{\beta}^{1/2}R_k\rho_{\beta}^{1/2}\|_2}{|{\rm tr}(\rho_{\beta}^{1/2}R^{\dagger}_k\rho_{\beta}^{1/2}R_k)|}.
\end{equation}
First, we upper bound the numerator of $\|L_k\|_2$.  Using Lemma 2 \eqref{Holder} for two times, the numerator is upper bounded as 
\begin{align}
    \|\rho_{\beta}^{1/2}R_k\rho_{\beta}^{1/2}\|_2 &=\|\rho_\beta^{1/4} \rho_\beta^{1/4} R_k \rho_\beta^{1/4}\rho_\beta^{1/4} \|_2 \le \|\rho_\beta^{1/4}\|_\infty \cdot \|\rho_\beta^{1/4} R_k \rho_\beta^{1/4}\rho_\beta^{1/4}\|_2 \nonumber\\
    &\le \|\rho_\beta^{1/4}\|_\infty^2 \cdot \|\rho_\beta^{1/4} R_k \rho_\beta^{1/4}\|_2=\sqrt{p_{\max}(\rho_{\beta}){\rm tr}(\rho_{\beta}^{1/2}R^{\dagger}_k\rho_{\beta}^{1/2}R_k)}.
\end{align}
The next step is to lower bound the denominator. We expand the denominator using the energy basis of the Gibbs state:
\begin{align}
    {\rm tr}(\rho_{\beta}^{1/2}R^{\dagger}_k\rho_{\beta}^{1/2}R_k)&=\sum_i\langle i|\rho_{\beta}^{1/2}R_k^{\dagger}\rho_{\beta}^{1/2}R_k|i\rangle=\sum_{i,j}\langle i|\rho_{\beta}^{1/2}R_k^{\dagger}|j\rangle\langle j|\rho_{\beta}^{1/2}R_k|i\rangle \\&=\sum_{i,j}\sqrt{p_ip_j}\langle i|R_k^{\dagger}|j\rangle\langle j|R_k|i\rangle \geq  p_{\min} \sum_{i,j}\langle i|R_k^{\dagger}|j\rangle\langle j|R_k|i\rangle= p_{\min}{\rm tr}(R_k^{\dagger}R_k)= p_{\min}\|R_k\|_2^2,
\end{align}
where in the first line we insert a $\sum_{j}|j\rangle\langle j|=\mathbb{I}$ and in the second line we use $\rho^{1/2}_{\beta}|i\rangle=\sqrt{p_i}|i\rangle$. Combining the upper bound for the numerator and the lower bound for the denominator, we finally obtain the desired upper bound:
\begin{equation}
    \|L_k\|_2\leq \frac{ \sqrt{p_{\max}(\rho_{\beta})}}{ \sqrt{{\rm tr}(\rho_{\beta}^{1/2}R^{\dagger}_k\rho_{\beta}^{1/2}R_k)}}\leq\frac{\sqrt{p_{\max}(\rho_{\beta})}}{\sqrt{p_{\min}(\rho_{\beta})}}\frac{1}{\|R_k\|_2}\Rightarrow O_k\leq \sqrt{\frac{ p_{\max}(\rho_{\beta})}{ p_{\min}(\rho_{\beta})}}.
\end{equation}
The upper bound for $\|L_k\|_2$ is also directly obtained as 
\begin{equation}
    \|L_k\|_2\leq \sqrt{\frac{ p_{\max}(\rho_{\beta})}{ p_{\min}(\rho_{\beta})}}\frac{1}{\|R_k\|_2}= \sqrt{\frac{ p_{\max}(\rho_{\beta})}{ p_{\min}(\rho_{\beta})}}\frac{\|R_k\|_1}{\|R_k\|_2}\leq \sqrt{\frac{ p_{\max}(\rho_{\beta})}{ p_{\min}(\rho_{\beta})}}\frac{\sqrt{d}\|R_k\|_2}{\|R_k\|_2}=\sqrt{d}\cdot\sqrt{\frac{ p_{\max}(\rho_{\beta})}{ p_{\min}(\rho_{\beta})}},
\end{equation}
where we use the normalization $\|R_k\|_1=1$.

\subsection{Degenerate eigenvalues}
We now show that the assumption of non-degenerate eigenvalues can be relaxed. Let $\lambda_k$ be an eigenvalue of $\mathcal L$ with multiplicity $m_k$ and
let $E_k$ be the corresponding eigenspace.

\textit{KMS-QDB Lindbladians}:
Because $\mathcal L$ is self-adjoint on the finite-dimensional inner-product space
$(\mathcal B(\mathcal H),\langle\cdot,\cdot\rangle_{\mathrm{KMS}})$,
there exists a KMS-orthonormal basis of right eigenoperators
$\{R_{k,\alpha}\}_{\alpha=1}^{m_k}\subset E_k$ such that
\begin{equation}
    \mathcal L(R_{k,\alpha})=\lambda_k R_{k,\alpha},\qquad
\langle R_{k,\alpha},R_{k,\beta}\rangle_{\mathrm{KMS}}=\delta_{\alpha\beta}.
\end{equation}
We define
\begin{equation}
    L_{k,\alpha}:=\frac{\mathcal G(R_{k,\alpha})}
{{\rm tr}\!\left(\rho_\beta^{1/2}R_{k,\alpha}^\dagger \rho_\beta^{1/2}R_{k,\alpha}\right)}.
\end{equation}
Then, using $\mathcal L^\dagger\!\circ\mathcal G=\mathcal G\!\circ\mathcal L$, one easily finds that
$\mathcal L^\dagger(L_{k,\alpha})=\lambda_k L_{k,\alpha}$ and
${\rm tr}(L_{k,\alpha}^\dagger R_{k,\beta})
=\langle R_{k,\alpha},R_{k,\beta}\rangle_{\mathrm{KMS}}=\delta_{\alpha\beta}$ hold. That is, $\{L_{k,\alpha}\}$ are still the biorthogonal left eigenoperators.

Following the same procedure, we get
$\|\rho_\beta^{1/2}R_{k,\alpha}\rho_\beta^{1/2}\|_2^2\le p_{\max}(\rho_\beta){\rm tr}(\rho_\beta^{1/2}R_{k,\alpha}^\dagger \rho_\beta^{1/2}R_{k,\alpha})$
and
${\rm tr}(\rho_\beta^{1/2}R_{k,\alpha}^\dagger \rho_\beta^{1/2}R_{k,\alpha})
\ge  p_{\min}(\rho_\beta)\|R_{k,\alpha}\|_2^2$.
Therefore, for each $\alpha$,
\begin{equation}
    \|L_{k,\alpha}\|_2
\le \sqrt{\frac{ p_{\max}(\rho_\beta)}{ p_{\min}(\rho_\beta)}}
\frac{1}{\|R_{k,\alpha}\|_2}
\quad\Longrightarrow\quad
\|L_{k,\alpha}\|_2\,\|R_{k,\alpha}\|_2
\le  \sqrt{\frac{ p_{\max}(\rho_\beta)}{ p_{\min}(\rho_\beta)}}.
\end{equation}

\textit{Davies generator:}

Since $\mathcal D$ is self-adjoint w.r.t.\ $\langle\cdot,\cdot\rangle_{\rho_\beta}$, it is diagonalizable;
because it commutes with $\mathcal U$, there exists a common eigenbasis
$\{R_{k,\alpha}\}_{\alpha=1}^{m_k}\subset E_k$ such that
\begin{equation}
    \mathcal D(R_{k,\alpha})=\nu_{k,\alpha}R_{k,\alpha},\qquad
\mathcal U(R_{k,\alpha})=u_{k,\alpha}R_{k,\alpha},
\end{equation}
and hence $\mathcal L(R_{k,\alpha})=(\nu_{k,\alpha}+iu_{k,\alpha})R_{k,\alpha}=:\lambda_{k,\alpha}R_{k,\alpha}$.
Moreover, using $\mathcal U^\dagger=-\mathcal U$, $[H,\rho_\beta]=0$ and
$\mathcal D^\dagger(X)=\mathcal D(X\rho_\beta^{-1})\rho_\beta$, we have $\mathcal L^\dagger(R_{k,\alpha}\rho_\beta)=\lambda_{k,\alpha}^*(R_{k,\alpha}\rho_\beta).$ Imposing the biorthonormality ${\rm tr}(L_{k,\alpha}^\dagger R_{k,\beta})=\delta_{\alpha\beta}$
fixes
\begin{equation}
    L_{k,\alpha}=\frac{R_{k,\alpha}\rho_\beta}{{\rm tr}(\rho_\beta R_{k,\alpha}^\dagger R_{k,\alpha})}.
\end{equation}
By H$\ddot{\rm o}$lder’s inequality and the spectral bounds of $\rho_\beta$,
$\|R_{k,\alpha}\rho_\beta\|^2_2\le p_{\max}(\rho_\beta){\rm tr}(\rho_\beta R_{k,\alpha}^\dagger R_{k,\alpha})$ and
${\rm tr}(\rho_\beta R_{k,\alpha}^\dagger R_{k,\alpha})\ge p_{\min}(\rho_\beta)\|R_{k,\alpha}\|_2^2$,
which yields, for every $\alpha$,
\[
\|L_{k,\alpha}\|_2\,\|R_{k,\alpha}\|_2
\le \sqrt{p_{\max}(\rho_\beta)/p_{\min}(\rho_\beta)}.
\]

Thus the upper bound obtained for the non-degenerate case holds verbatim for each vector in a degenerate eigenspace.

\section{Condition for the typical strong (Quantum) Mpemba effect}
In this section, we discuss the sufficient condition for the occurrence of the typical strong Quantum Mpemba effect proposed in the main text.
Following the spirit of \cite{Mori20resolving}, we introduce the (typical) $\epsilon$-relaxation time corresponding to the $k$th eigenmode ($k\geq2$) via the relation
\begin{equation}
    |\langle a_k\rangle| e^{-|{\rm Re}(\lambda_k)|t_{\epsilon,k}}=\epsilon,
\end{equation}
which gives 
\begin{equation}
    t_{\epsilon,k}:=\frac{1}{|{\rm Re}(\lambda_k)|}\log{\left(\frac{|\langle a_k\rangle|}{\epsilon}\right)}.
\end{equation}
The (typical) relaxation time is roughly given by $t_{\epsilon}=\max_kt_{\epsilon,k}$. For many cases, $t_{\epsilon}=t_{\epsilon,2}$ because the spectral gap $|{\rm Re}(\lambda_2)|$ is the smallest among all $|{\rm Re}(\lambda_k)|,\ k\geq2$. This is why it is commonly believed that the relaxation time is determined by the spectral gap as $\tau_{\rm rel}\sim1/|{\rm Re}(\lambda_2)|$. 

However, ``faster'' ($k>2$) modes may be dominant when the system-size effect is taken into account in high-dimensional open quantum systems (e.g., many-body systems). Consider cases where both the second mode and the $k$th mode concentrate to their mean value. There are some systems, where $\lim_{d\to \infty}\langle a_2\rangle={\rm tr}(L_2^{\dagger})/d\to 0$ and $\lim_{d\to \infty}\langle a_k\rangle>0$. An example is the Davies map, if $\lambda_2$ is complex, then ${\rm tr}(L_2^{\dagger})=0\Rightarrow \langle a_2\rangle=0$ for all $d$. There must be some real eigenvalues $\lambda_k$, and the corresponding $\langle a_k\rangle >0$. In this case, if ${\rm Var}(a_2)\sim \mathcal{O}(d^{-2a})$, then $|\langle a_2\rangle|\sim \mathcal{O}(d^{-a})$.

Now, let us compare the dimension scaling of $t_{\epsilon,2}$ and $t_{\epsilon,k}$. To make their definitions physical (non-negative), we choose the threshold as $\epsilon=d^{-a}\epsilon_0$, where $\epsilon_0$ is a small constant independent of the system dimension. With this choice, we have
\begin{align}
    t_{\epsilon,2}&\sim\frac{1}{|{\rm Re}(\lambda_2)|}\log{\left(\frac{1}{\epsilon_0}\right)}\sim \frac{\mathcal{O}(1)}{|{\rm Re}(\lambda_2)|},\\
    t_{\epsilon,k}&\sim\frac{1}{|{\rm Re}(\lambda_k)|}\log{\left(\frac{d^a}{\epsilon_0}\right)}\sim \frac{\mathcal{O}(\log d)}{|{\rm Re}(\lambda_k)|}.
\end{align}
Therefore, if the condition 
\begin{equation}
    \frac{|{\rm Re}(\lambda_k)|}{|{\rm Re}(\lambda_2)|}\lesssim\mathcal{O}(\log d)
\end{equation}
is satisfied, the long-time relaxation will be dominated by the $k$th mode instead of the slowest mode. That is, the relaxation timescale would be more appropriately characterized by $1/|{\rm Re}(\lambda_k)|$ instead of $1/|{\rm Re}(\lambda_2)|$ in this case, and this is the onset of the typical strong quantum Mpemba effect. From the above analysis, it is clear that this effect can occur even when there is Liouvillian gap closing (the only requirement is that the gap closing speed is not too fast compared to the $k$th mode).

Notably, the requirement $\lim_{d\to \infty}\langle a_k\rangle>0$ can be relaxed to $\langle a_k\rangle$ has a smaller $d$-scaling than $\langle a_2\rangle$. 

\section{Upper bound for ${\rm Var}(a_k)$ under approximate unitary 2-designs}

In all three sampling schemes considered in the main text and in this SM (random pure state, random mixed state from an induced/HS measure, and a random state from a constrained subspace $\mathcal H_R$), the variance of $a_k$ depends only on the first and second moments of the unitary ensemble used to generate $\rho$ (or $\rho_S$). It is therefore natural to quantify deviations from Haar by the deviation of the \emph{twirl channels}. Let $\nu$ be a distribution over unitaries on this $D$-dimensional space. Define the one- and two-fold twirls
\begin{equation}
\mathcal T^{(1)}_\nu(X):=\mathbb E_{U\sim\nu}\,U X U^\dagger,\qquad
\mathcal T^{(2)}_\nu(Y):=\mathbb E_{U\sim\nu}\,U^{\otimes2} Y (U^\dagger)^{\otimes2}.
\label{eqS:twirl-def}
\end{equation}
We call $\nu$ an $\varepsilon$-approximate unitary $2$-design (diamond version) if
\begin{equation}
\big\| \mathcal T^{(2)}_\nu - \mathcal T^{(2)}_{\rm Haar} \big\|_\diamond:=\max_{\rho}\|\mathcal T^{(2)}_\nu(\rho) - \mathcal T^{(2)}_{\rm Haar}(\rho)\|_1 \le \varepsilon .
\label{eqS:approx2-diamond}
\end{equation}

Fix an arbitrary unit vector $|\phi_0\rangle$ in the underlying space (for the constrained case $|\phi_0\rangle\in\mathcal H_R$). Sampling the random pure state as $\rho=U|\phi_0\rangle\langle\phi_0|U^\dagger$ with $U\sim\nu$ gives
\begin{equation}
\mathbb E_\nu\big[\rho^{\otimes2}\big]
=\mathbb E_{U\sim\nu}\,U^{\otimes2}\big(|\phi_0\phi_0\rangle\!\langle\phi_0\phi_0|\big)(U^\dagger)^{\otimes2}
=\mathcal T^{(2)}_\nu(Y_0),
\qquad
Y_0:=|\phi_0\phi_0\rangle\!\langle\phi_0\phi_0|.
\label{eqS:rho2-as-twirl}
\end{equation}
Note that $Y_0\ge0$, ${\rm tr}\,Y_0=1$, $\|Y_0\|_1=1$, and $\|Y_0\|_2=1$.

For an operator $L_k^\dagger$ on $\mathcal H_S$ and the reduced state $\rho_S$, we consider
\[
a_k={\rm tr}_S(L_k^\dagger\rho_S)={\rm tr}\big[(L_k^\dagger\!\otimes\!\mathbb I)\rho\big],
\qquad
|a_k|^2={\rm tr}\big[(L_k^\dagger\!\otimes\!\mathbb I\!\otimes\!L_k\!\otimes\!\mathbb I)\,\rho^{\otimes2}\big].
\]
Introduce the fixed observable on the doubled space
\begin{equation}
W_k:=L_k^\dagger\!\otimes\!L_k\!\otimes\!\mathbb I,
\qquad \|W_k\|_\infty=\|L_k\|_\infty^2,\quad \|W_k\|_2=\|L_k\|_2^2.
\label{eqS:Wk-def}
\end{equation}
Combining \eqref{eqS:rho2-as-twirl} with the above,
\begin{equation}
\mathbb E_\nu|a_k|^2
={\rm tr}\!\big[W_k\,\mathcal T^{(2)}_\nu(Y_0)\big],\qquad
\mathbb E_{\rm Haar}|a_k|^2
={\rm tr}\!\big[W_k\,\mathcal T^{(2)}_{\rm Haar}(Y_0)\big].
\label{eqS:second-moment-functional}
\end{equation}

The error on variance can be upper bounded as
\begin{equation}
\Delta{\rm Var}:=\big|{\rm Var}_\nu(a_k)-{\rm Var}_{\rm Haar}(a_k)\big|
\ \le\ \underbrace{\big|\mathbb E_\nu|a_k|^2-\mathbb E_{\rm Haar}|a_k|^2\big|}_{\text{second moment}}
\ +\ \underbrace{\big|\,|\mathbb E_\nu a_k|^2-|\mathbb E_{\rm Haar} a_k|^2\,\big|}_{\text{first moment}},
\label{eqS:split}
\end{equation}
where ${\rm Var}_{\rm Haar}(a_k)={\rm Var}(a_k)$ is the variance considered in the main text, by definition of the unitary 2-design. 

First the second moment term, by H$\ddot{\rm o}$lder and the definition of diamond norm,
\begin{align}
\big|\mathbb E_\nu|a_k|^2-\mathbb E_{\rm Haar}|a_k|^2\big|
&=\big|{\rm tr}\big[W_k\,(\mathcal T^{(2)}_\nu-\mathcal T^{(2)}_{\rm Haar})(Y_0)\big]\big|
\le \|W_k\|_\infty\,\big\|(\mathcal T^{(2)}_\nu-\mathcal T^{(2)}_{\rm Haar})(Y_0)\big\|_1
\notag\\
&\le \|W_k\|_\infty\,\big\|\mathcal T^{(2)}_\nu-\mathcal T^{(2)}_{\rm Haar}\big\|_\diamond\,\|Y_0\|_1
\le \varepsilon\,\|L_k\|_\infty^2.
\label{eqS:second-diamond}
\end{align}
Here we used that for any linear map $\Phi$, $\|\Phi(Z)\|_1\le\|\Phi\|_\diamond\|Z\|_1$ (the diamond norm dominates the induced trace norm).

For the first moment term, we first consider a relation: for any choice of model, there exists a CPTP map $\mathcal J$ of the form (copying the Hilbert space)
\begin{equation}
\mathcal J(X)=X\otimes\tau,\qquad \tau\ge0,\ {\rm tr}\,\tau=1,
\label{eqS:J-def}
\end{equation}
such that
\begin{equation}
\mathcal T^{(1)}=\operatorname{tr}_2\circ \mathcal T^{(2)}\circ\mathcal J,
\label{eqS:factorization}
\end{equation}
where the trace ${\rm tr}_2$ is over the copied Hilbert space.

\emph{Proof.} For every unitary $U$,
$\operatorname{tr}_2\!\big[(U\!\otimes\!U)(X\!\otimes\!\tau)(U^\dagger\!\otimes\!U^\dagger)\big]
=UXU^\dagger\,{\rm tr}(U\tau U^\dagger)=UXU^\dagger$; averaging over $\nu$ yields \eqref{eqS:factorization}. $\square$

Then, \eqref{eqS:factorization} and submultiplicativity of the diamond norm imply
\[
\big\|\mathcal T^{(1)}_\nu-\mathcal T^{(1)}_{\rm Haar}\big\|_\diamond
\le \|\operatorname{tr}_2\|_\diamond\,\|\mathcal T^{(2)}_\nu-\mathcal T^{(2)}_{\rm Haar}\|_\diamond\,\|\mathcal J\|_\diamond
\le \varepsilon,
\]
because CPTP maps have diamond norm $1$. Hence, for any fixed reference state $\rho_{0,S}$ (density matrix),
\begin{equation}
\big|\mathbb E_\nu a_k-\mathbb E_{\rm Haar} a_k\big|
=\big|{\rm tr}\!\big(L_k^\dagger\,[\mathcal T^{(1)}_\nu-\mathcal T^{(1)}_{\rm Haar}](\rho_{0,S})\big)\big|
\le \|L_k\|_\infty\,\varepsilon.
\label{eqS:first-mean-diamond}
\end{equation}
Finally, since $|{\rm tr}(L_k^\dagger \sigma)|\le \|L_k\|_\infty$ for any state $\sigma$,
\begin{equation}
\big|\,|\mathbb E_\nu a_k|^2-|\mathbb E_{\rm Haar} a_k|^2\,\big|
=|\mathbb E_\nu a_k+\mathbb E_{\rm Haar} a_k||\mathbb E_\nu a_k-\mathbb E_{\rm Haar} a_k|\le 2\,\|L_k\|_\infty\,\big|\mathbb E_\nu a_k-\mathbb E_{\rm Haar} a_k\big|
\le 2\,\varepsilon\,\|L_k\|_\infty^2.
\label{eqS:first-square-diamond}
\end{equation}
Combining \eqref{eqS:second-diamond} and \eqref{eqS:first-square-diamond} we obtain
\begin{equation}
\big|{\rm Var}_\nu(a_k)-{\rm Var}_{\rm Haar}(a_k)\big|
\le 3\,\varepsilon\,\|L_k\|_\infty^2.
\label{eqS:diamond-variance-bound}
\end{equation}


\section{Derivation of the perturbation bound $O_k\|\delta \mathcal{L}\|_{\rm op}$ mentioned in the footnote}

In this section, we provide a detailed derivation for the first-order correction to an eigenvalue of the Lindbladian $\mathcal{L}$ under a small perturbation. 

Let $\lambda_k$ be a non-degenerate eigenvalue of $\mathcal{L}$ with corresponding right and left eigenvectors, $R_k$ and $L_k$, respectively. They satisfy the eigenvalue equations:
\begin{align}
    \mathcal{L} R_k &= \lambda_k R_k, \label{eq:right_eig} \\
    \mathcal{L}^\dagger L_k &= \lambda_k^* L_k. \label{eq:left_eig}
\end{align}
Now, we introduce a small perturbation $\delta\mathcal{L}$ to the Lindbladian, such that the new operator is $\mathcal{L}' = \mathcal{L} + \delta\mathcal{L}$. The new eigenvalue $\lambda'_k$ and right eigenvector $R'_k$ can be expanded to first order as:
\begin{align}
    \lambda'_k &= \lambda_k + \delta\lambda_k + O((\delta\mathcal{L})^2), \\
    R'_k &= R_k + \delta R_k + O((\delta\mathcal{L})^2),
\end{align}
where $\delta\lambda_k$ and $\delta R_k$ are the first-order corrections. The new eigenvalue equation is $(\mathcal{L} + \delta\mathcal{L})(R_k + \delta R_k) = (\lambda_k + \delta\lambda_k)(R_k + \delta R_k)$. Expanding this equation and keeping only terms up to the first order in the perturbation yields:
\begin{equation}
    \mathcal{L}R_k + \mathcal{L}\delta R_k + \delta\mathcal{L}R_k \approx \lambda_k R_k + \lambda_k \delta R_k + \delta\lambda_k R_k.
\end{equation}
The zeroth-order terms, $\mathcal{L}R_k = \lambda_k R_k$, cancel out, leaving the equation for the first-order corrections:
\begin{equation}
    (\mathcal{L} - \lambda_k \mathcal{I}) \delta R_k = \delta\lambda_k R_k - \delta\mathcal{L}R_k,
    \label{eq:first_order_eq}
\end{equation}
where $\mathcal{I}$ is the identity super-operator.

To solve for the eigenvalue correction $\delta\lambda_k$, we project this equation onto the corresponding left eigenvector $L_k$. This is achieved by left-multiplying by $L_k^\dagger$ and taking the trace over the operator space. The left-hand side of Eq.~\eqref{eq:first_order_eq} becomes:
\begin{align}
    {\rm tr}\left[ L_k^\dagger (\mathcal{L} - \lambda_k \mathcal{I}) \delta R_k \right] &= {\rm tr}(L_k^\dagger \mathcal{L} \delta R_k) - \lambda_k {\rm tr}(L_k^\dagger \delta R_k) \nonumber \\
    &= {\rm tr}[(\mathcal{L}^\dagger L_k)^\dagger \delta R_k] - \lambda_k {\rm tr}(L_k^\dagger \delta R_k) \nonumber \\
    &= {\rm tr}[(\lambda_k^* L_k)^\dagger \delta R_k] - \lambda_k {\rm tr}(L_k^\dagger \delta R_k) \nonumber \\
    &= \lambda_k {\rm tr}(L_k^\dagger \delta R_k) - \lambda_k {\rm tr}(L_k^\dagger \delta R_k) = 0.
\end{align}
Here, we used the definition of the adjoint operator in the second line and the left eigenvector equation, Eq.~\eqref{eq:left_eig}, in the third line. Since the left-hand side vanishes, the right-hand side must also be zero:
\begin{equation}
    {\rm tr}\left[ L_k^\dagger (\delta\lambda_k R_k - \delta\mathcal{L}R_k) \right] = 0.
\end{equation}
By rearranging the terms, we solve for the first-order correction to the eigenvalue:
\begin{equation}
    \delta\lambda_k = \frac{{\rm tr}(L_k^\dagger \delta\mathcal{L} R_k)}{{\rm tr}(L_k^\dagger R_k)}={\rm tr}(L_k^\dagger \delta\mathcal{L} R_k),
    \label{eq:perturb_final}
\end{equation}
where the normalization ${\rm tr}(L_k^\dagger R_k)=1$ has been used.

Next, we derive an upper bound on the magnitude of this correction. The $|\delta \lambda_k|$ can be bounded using the Cauchy-Schwarz inequality for the HS inner product, which states $|{\rm tr}(A^\dagger B)| \le \|A\|_{2} \|B\|_{2}$. We set $A=L_k$ and $B=\delta\mathcal{L} R_k$ and obtain
\begin{equation}
   |\delta \lambda_k|= |{\rm tr}(L_k^\dagger (\delta\mathcal{L} R_k))| \le \|L_k\|_{2} \|\delta\mathcal{L} R_k\|_{2}.
\end{equation}
The second term on the right-hand side can be further bounded as
\begin{equation}
    \|\delta\mathcal{L} R_k\|_{2} \le \|\delta\mathcal{L}\|_{
    \rm op} \|R_k\|_{2}.
\end{equation}
Combining these inequalities, we obtain the desired upper bound for the eigenvalue sensitivity:
\begin{equation}
    |\delta\lambda_k| \le O_k \|\delta\mathcal{L}\|_{\rm op}.
\end{equation}
This bound shows that the overlap $O_k$ acts as the condition number for the eigenvalue $\lambda_k$, quantifying the degree of amplification of the perturbation $\delta\mathcal{L}$.

\section{The consequence of using different normalization}
If we normalize the right eigenvectors using the HS norm as in \cite{25randomLindblad}, that is, $\|R_k\|_2=1$ for $k\geq 2$, the scaling condition for the concentration of the $k$th mode will be much weaker than its current form in the main text ($O_k$ grows slower than $\mathcal{O}(d^{1/2})$ for random pure initial states and slower than $\mathcal{O}(d^{1})$ for random mixed initial states). In this case, $O_k=\|L_k\|_2\|R_k\|_2=\|L_k\|_2$, so that the upper bounds of the variance become
\begin{equation}
    {\rm Var}_U(a_k)\leq \frac{O_k^2}{d^2}, \quad {\rm Var}(a_k)\leq \frac{O_k^2}{d^3}
\end{equation}
for initial random pure states and random mixed states, respectively.
 \begin{figure}
    \centering
     \includegraphics[width=0.9\linewidth]{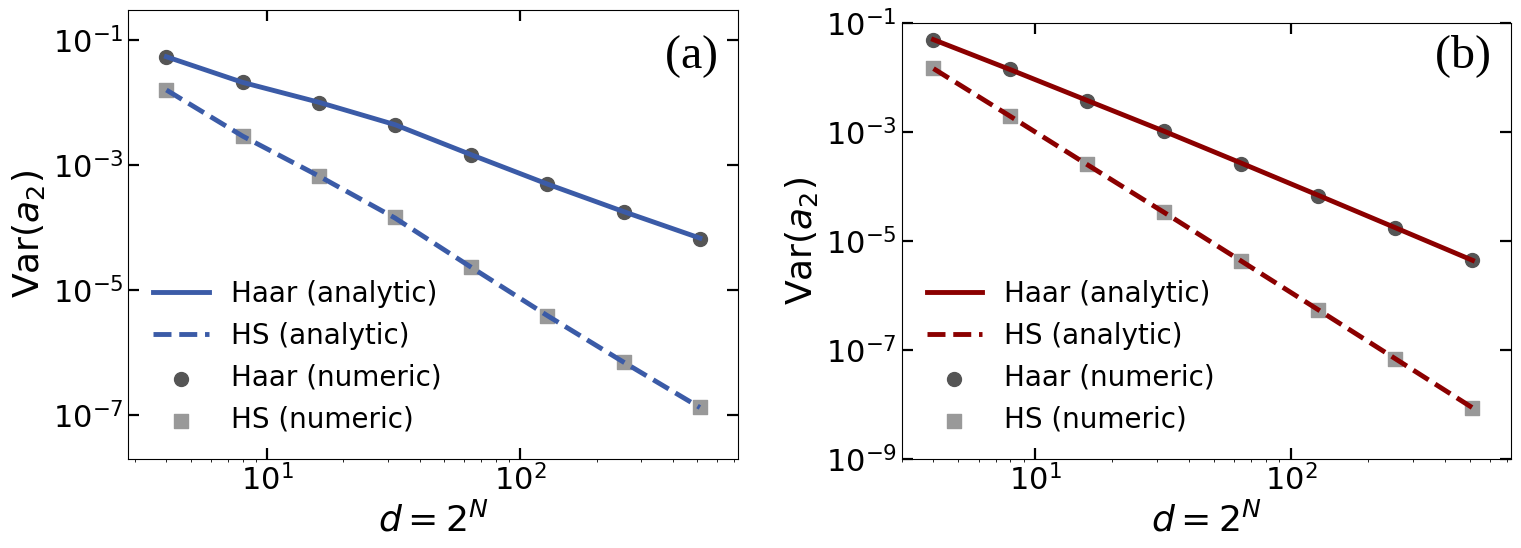}
     \caption{Variance of $a_2$ for the dissipative TFIM with different spin number $N$. The normalization for the right eigenvectors is $\| R_k\|_2=1,\ k\geq2$. The system size varies from $N=2$ to $N=9$ ($d=512$). Other parameters are $J=g=1.0$, $\gamma=0.5$. (a) $\beta=0.1$, high temperature (b) $\beta=100$, low temperature.}
    \label{S1}
\end{figure}
It is obvious that the convergence speed under this normalization is much larger, e.g., see Fig. \ref{S1}.

We can use this normalization when the relaxation to stationarity is characterized by the HS norm $\|\rho(t)-\rho_{\rm ss}\|_2$, as in Ref. \cite{21prl_QMpemba}. However, the use of the HS norm instead of the trace norm implies that one has a relatively low standard for distinguishability, given that the trace norm is larger than the HS norm.

\bibliography{refs}